\documentclass[5p,twocolumn,number]{elsarticle}
\usepackage{array,amsmath,amssymb,booktabs,graphicx,hyperref,microtype}
\usepackage[T1]{fontenc}\usepackage{lmodern}
\hypersetup{colorlinks=true,citecolor=blue,linkcolor=blue,urlcolor=blue}
\newcommand{\Msun}{M_\odot}
\newcommand{\Ycf}{Y_{254}}
\newcommand{\cf}{^{254}\mathrm{Cf}}\newcommand{\qth}{\dot q_{\rm th}}
\AtBeginDocument{\let\bfseries\mdseries\let\itshape\upshape}

\newcommand{\colhead}[1]{\begin{tabular}[c]{@{}c@{}}#1\end{tabular}}
\journal{Journal of High Energy Astrophysics}
\newcommand{\Yatot}{Y_{\alpha,\mathrm{tot}}}
\newcommand{\Yatrue}{Y_{\alpha,\mathrm{true}}}
\newcommand{\Yacap}{Y_{\alpha,\mathrm{cap}}}
\newcommand{\Yazero}{Y_{\alpha,0}}
\begin{document}\begin{frontmatter}
\title{A Unified Model for Radioactive Signatures and Source Identifiability in Neutron Star Mergers}
\author[z,c]{Hisham Anwer}
\ead{hisham@cantab.net}
\affiliation[z]{organization={Physics of the Universe Program, Zewail City of Science and Technology},city={Giza},country={Egypt}}
\affiliation[c]{organization={Department of Physics, Faculty of Science, Cairo University},city={Giza},country={Egypt}}

\begin{abstract}
Late light curves of neutron star mergers can contain substantial fission power while remaining compatible with different radioactive inventories. We develop a unified source attribution model in which a common bounded response formulation links nuclear heating, radiation transport and isotope identification. Within particle dependent deposition and gray photon escape, matched luminosity comparisons show how long alpha and fission responses become aligned. Explicit competing inventories then reproduce a light curve with a substantial $^{254}$Cf contribution, with smooth evolution of the beta luminosity response providing an important source of compensation. We derive optimization certificates and collective likelihood bounds that quantify the resulting limits on identification under specified source and ejecta constraints. We also introduce an energy conserving model with two radiation frequency groups to resolve the information carried by color at fixed total luminosity precision. Its exact response modes identify the source adjustments that absorb color contrast, while a residual color floor characterizes the contrast retained by the permitted family. Late observations of AT~2023vfi with the James Webb and Hubble Space Telescopes provide an observational benchmark: the reference and retained alternatives predict bolometric declines about 1.5 quoted uncertainties shallower than the reconstruction, while differences between inventories containing and excluding $^{254}$Cf remain within the reported decline uncertainty. The unified model provides a quantitative basis for connecting nucleosynthesis predictions, gravitational wave constraints and the temporal and spectral information in merger observations to radioactive source attribution.
\end{abstract}
\end{frontmatter}
\raggedbottom

\section{Introduction}
Gravitational wave and kilonova observations of GW170817 established a direct connection between neutron star mergers and heavy element production \cite{Abbott2017,Kasen2017,Metzger2020}. At late times, the emission probes the radioactive inventory of the expanding ejecta. As the broad beta decay background fades, individual alpha chains and fissioning nuclei can contribute appreciably to the heating. The $^{223}$Ra and precursor-fed $^{225}$Ra--$^{225}$Ac chains and $\cf$ have therefore attracted attention as potential radioactive fingerprints \cite{Wu2019,Zhu2018,Chen2024}. Their observable signatures are shaped by particle deposition, energy storage and photon escape, which connect nuclear decay power to the emerging light curve \cite{Barnes2016,KasenBarnes2019,Hotokezaka2020}.

The relevant late time epochs are now accessible to observation. JWST followed AT~2023vfi, the transient associated with GRB~230307A, at approximately 29 and 61 observer-frame days; its spectral interpretation illustrates the importance of atomic data and radiation transport \cite{Levan2024,Gillanders2025}. Network calculations further show that trajectories with similar electron fraction can produce markedly different actinide-rich spectra and light curves \cite{Pognan2025}. Previous studies have established the heating contributions of particular nuclei and the dependence of merger emission on nuclear inputs, thermalization and opacity \cite{Wu2019,Zhu2018,Chen2024,Zhu2021,Bulla2023,Anwer2026}. These developments sharpen the question of source identification: recovering an isotope requires its radiative response to remain distinguishable from the responses of admissible competing sources and ejecta configurations.

A large luminosity fraction and an identifiable nuclear signature are different physical properties. We compare calculations at the same late fission share to isolate how deposition changes source distinguishability. Bounded beta and alpha mixtures reproduce a Cf-bearing light curve under these matched conditions, extending radioactive fingerprint studies \cite{Wu2019,Zhu2018} from source contributions to attribution and connecting them to thermalization and opacity uncertainties \cite{Bulla2023}. A source separation map independently varies the injected abundance and competing yield budget to resolve the role of nuclear restrictions.

\begin{figure*}[t]\centering
\includegraphics[width=0.485\textwidth]{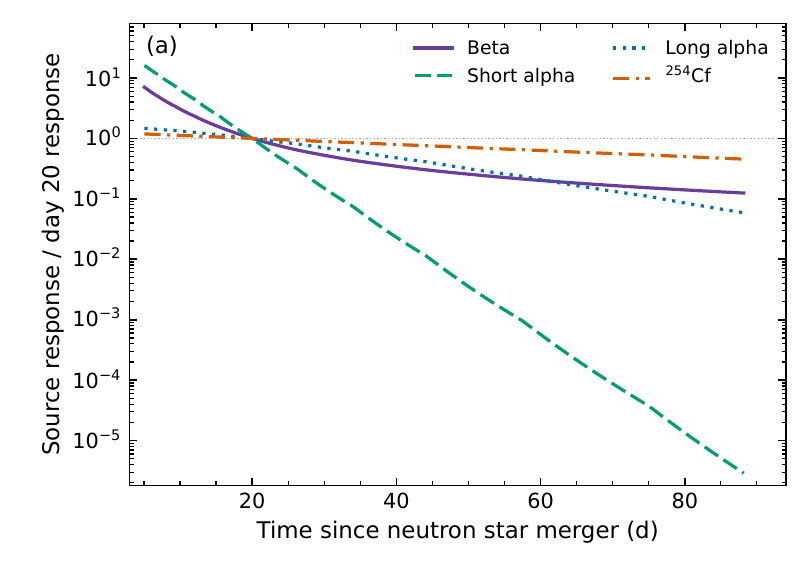}\hfill\includegraphics[width=0.485\textwidth]{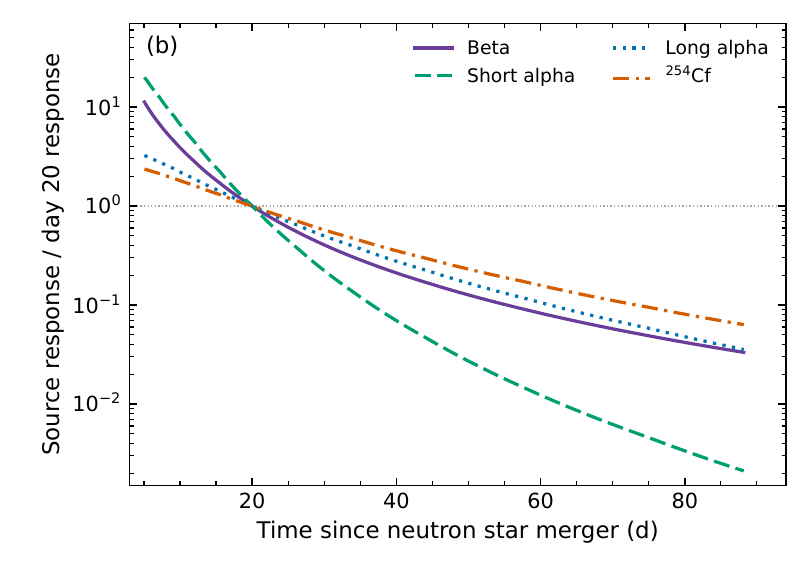}\\[2mm]\includegraphics[width=0.485\textwidth]{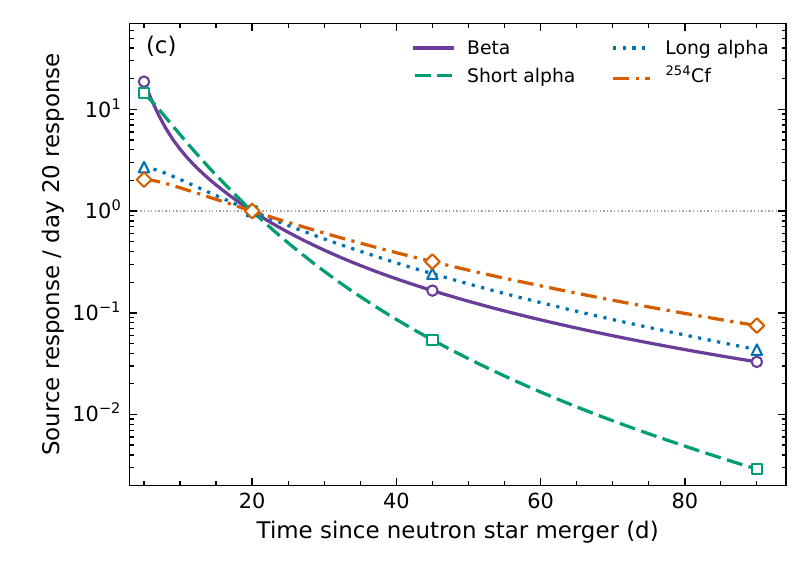}\hfill\includegraphics[width=0.485\textwidth]{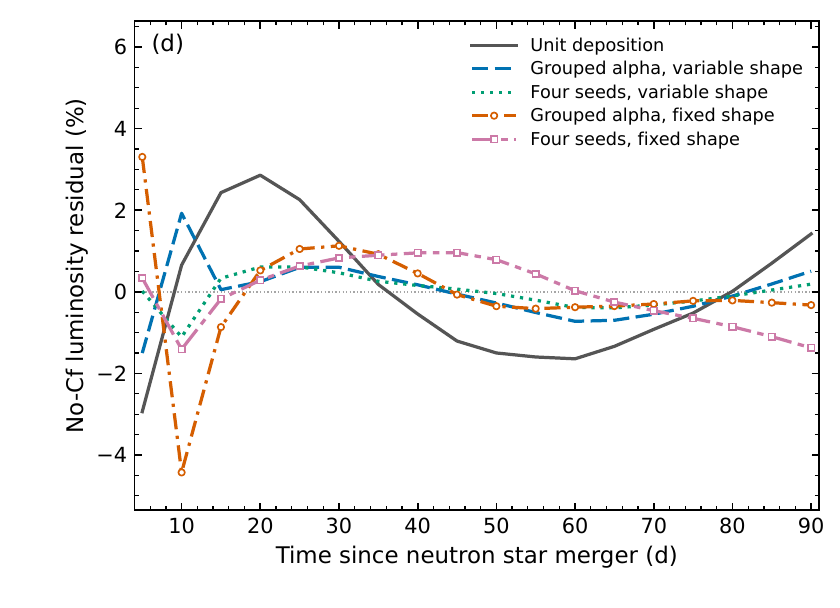}
\caption{Deposition and escape reshape source attribution. Beta, short alpha ($A=222,224$), long alpha ($A=223,225$) and Cf responses are normalized at day 20 in (a) emitted depositable power, (b) HN deposited power and (c) gray luminosity. Horizontal guides fix normalization; open symbols in (c) reproduce scalar-solver values. Long alpha and Cf become increasingly similar. (d) Feasible no-Cf residuals compare matched-share unit deposition with grouped and independent HN alpha sources at variable and fixed beta shape. Open symbols mark fixed-shape solutions. Straight segments connect evaluated epochs.}\label{fig:master}\end{figure*}

Our unified formulation applies one bounded family of transported inventories to luminosity comparisons, likelihood limits and color information. Established decay, particle deposition and gray escape prescriptions supply the physical baseline. We derive optimization certificates and collective likelihood bounds, then introduce an energy conserving extension with two radiation frequency groups. Its exact response modes resolve which source adjustments absorb the color contrast. This connects radioactive heating and spectral predictions \cite{Wu2019,Zhu2018,Pognan2025} to quantitative conditions for isotope identification. Figure~\ref{fig:master} displays the underlying decay-to-radiation mechanism.

\section{Nuclear heating and radiative response}\label{sec:transport}
\subsection{Decay, deposition and photon escape}
Radioactive decay supplies energy to merger ejecta, while particle energy loss and the radiation reservoir determine how that energy reaches the observer. The forward model follows each source through decay, deposition and photon escape so that changes in nuclear power can be distinguished from changes in its observable response. Time $t$ is measured from the controlled merger initialization, and number abundances $Y_c$ count initial seed nuclei per baryon; $m_u$ is the atomic mass unit. A chain seed identifies its initial nuclide and all descendants. The four alpha-chain seeds are $^{222}$At, $^{223}$At, $^{224}$Rn and $^{225}$Rn. With half-life $T_{1/2,i}$, decay rate $\lambda_i=\ln2/T_{1/2,i}$ and branch probabilities $b_{i\to j}$, the chain population satisfies $\vec n(t)=e^{Bt}\vec n(0)$, $B_{ii}=-\lambda_i$ and $B_{ji}=b_{i\to j}\lambda_i$. For particle class $p$,
\begin{equation}
 \begin{aligned}
 \dot q_p(t)&=m_u^{-1}\sum_cY_c\sum_{i\in c}\lambda_i n_{i|c}(t)\bar E_{i,p},\\
 \qth&=\sum_p\mathcal D_p[\dot q_p](t;M,v).
 \end{aligned}
 \label{eq:heating}
\end{equation}
Here $c$ indexes seeds, $i$ indexes chain nuclides, and $n_{i|c}$ is the population per initial seed ($n_{c|c}(0)=1$, with other populations initially zero). The matrix $B$ is the decay generator; $\bar E_{i,p}$ is the branch averaged energy per decay in particle class $p$. Thus $\dot q_p$ and $\qth$ are specific powers in erg s$^{-1}$ g$^{-1}$; neutrino energy is excluded. In velocity expressions, $c$ denotes the speed of light. Here $\mathcal D_p$ is the time dependent deposition operation, including previously injected charged particles; its output need not equal an instantaneous efficiency times the current input. The adopted prescription supplies particle specific electron, gamma, alpha and fission fragment deposition \cite{Hotokezaka2020}. The two stage $^{225}$Ra--$^{225}$Ac activity peaks at
\begin{equation}
 t_{\rm act,pk}=\frac{\ln(\lambda_{\rm Ac}/\lambda_{\rm Ra})}{\lambda_{\rm Ac}-\lambda_{\rm Ra}}=17.49\,\mathrm d,
 \label{eq:clock}
\end{equation}
when no daughter is present initially. This epoch specifies the nuclear activity maximum; deposition and photon escape determine the associated photometric response. The full network differs from the two stage approximation by at most $0.82\%$ after day one on the evaluated grid.

We calculate the beta heating background following the prescriptions provided in~\cite{Hotokezaka2020}, using a solar r-process residual abundance pattern over $88\leq A\leq209$. The reference mass weights are proportional to $A Y_A^\odot$ and normalized over this interval, where $Y_A^\odot$ denotes the solar residual number abundance at mass number $A$. The decay network uses the adopted nuclear half-lives, branching probabilities and branch averaged electron, gamma and alpha energies as physical inputs; the heating histories follow from the evolving populations in Eq.~(\ref{eq:heating}). For $^{254}$Cf, the mean lifetime is $7.541255\times10^6$ s, corresponding to a half-life of 60.5 d. We represent its depositable fission energy by two effective charged fragments with initial kinetic energies of 79.5 and 102.1 MeV, totaling 181.6 MeV per decay, and follow their time dependent energy loss using the particle deposition prescription \cite{Hotokezaka2020}. Neutrino energy is excluded.

For each independently deposited source, gray shell energies evolve as \cite{Hotokezaka2020}
\begin{equation}
 \dot E_s=M_s\qth-E_s/t-L_s,\quad
 L_s=\frac{\operatorname{erfc}\sqrt{t_{{\rm dif},s}/(2t)}}{\min(t,t_{{\rm dif},s})+\beta_st}E_s.
 \label{eq:transport}
\end{equation}
Here $s$ indexes radial shells: $M_s$, $E_s$ and $L_s$ are shell mass, stored radiation energy and escaping luminosity. The dimensionless shell speed is $\beta_s=v_s/c$, and $t_{{\rm dif},s}(t)=\tau_s(t)\beta_st$ is the time dependent diffusion time, with outward optical depth $\tau_s(t)=\int_{v_st}^{v_{\max}t}\kappa(r,t)\rho(r,t)\,dr$. The function $\operatorname{erfc}$ is the complementary error function. Homologous expansion means $r=v_st$; $\rho(v,t)\propto t^{-3}v^{-4.5}$ is normalized to the total mass. Photon integration starts at 0.01 d with zero stored radiation energy.

The baseline has $M=0.023\Msun$ ($\Msun$ is the solar mass), $v_{\min}=0.1c$, $v_{\max}=0.4c$, $\rho\propto v^{-4.5}$, and opacities $\kappa_{\rm in}=3$ and $\kappa_{\rm out}=0.5\,\mathrm{cm^2\,g^{-1}}$, separated at $0.14c$. Changing mass or velocity requires recomputing particle deposition before photon transport. At fixed ejecta, the shell equations are linear and the separately calculated source luminosities add.

We distinguish the injected total alpha number abundance $\Yatrue{}$ from the competing budget $\Yacap{}$. The reference $\Yazero{}=1.35\times10^{-4}$ normalizes the source kernels and supplies the local and moment diagnostics; the principal bounded comparison uses $\Yatrue{}=\Yacap{}=10^{-5}$. Isotope labels such as $^{225}$Ra denote mass number.

Figure~\ref{fig:master}(a--c) follows the same sources from nuclear power to photons, normalized at day 20. Particle energy retention and radiation storage alter their relative evolution before escape.

Independent solutions quantify the numerical consistency of the adopted prescriptions: the raw alpha network agrees with the reference decay calculation within $0.032\%$, and gray luminosities differ by at most $1.65\%$ at 2--29 d. Refining 120 shells and a 0.015 d step to 200 shells and a 0.005 d step changes the selected 5--90 d luminosities by at most $0.0122\%$, well below the stipulated 10\% flux errors.

\subsection{Thermalization and response geometry}
Particle deposition retains the history of injected decay energy. Consequently, this memory changes the temporal response of a source even when its nuclear evolution is fixed. In general, $q_{j,\rm dep}(t)=\sum_p\int_0^t K_{jp}(t,t';M,v)q_{jp,\rm raw}(t')\,dt'$, followed by the linear gray transport operator $\mathcal T_{M,v,\kappa}$. For fixed ejecta and a specified injection energy distribution, $K_{jp}$ denotes the linear causal response of $\mathcal D_p$ to a separately seeded history; it is operator notation rather than an independently tabulated kernel. Energy storage and delayed loss mean that $q_{j,\rm dep}(t)/q_{j,\rm raw}(t)$ measures an instantaneous power ratio. For example, for $A=222$, this ratio reaches 13.2 near 60 d as stored particle energy continues to deposit after the contemporaneous source has faded. Cumulative energy accounting, through Eq.~(\ref{eq:katz}), connects this delayed deposition to the radiation reservoir.

A source response is its emitted power, deposited power, or transported luminosity as a function of time. Below, HN denotes the adopted particle dependent deposition prescription; unit deposition sets the deposited power equal to the contemporaneous depositable power. Let $u_j^{(s)}(t_i)$ be the source contribution at stage $s$ divided by the \textrm{same stage total} at 18 epochs $t_i=5,10,\ldots,90$ d. We group $A=222,224$ as $S$ and $A=223,225$ as $H$, retain $F=\cf$, and use total alpha seed abundance $\Yatot=\sum_{j=222}^{225}Y_j=10^{-5}$ and Cf abundance $C=Y_{254}=2\times10^{-6}$. Here $A$ denotes mass number, while $\Yatot$ denotes the total alpha seed abundance. Stage superscript $s$ is separate from the shell index in Eq.~(\ref{eq:transport}). For normalized columns $\hat u_j=u_j/\|u_j\|_2$, define
\begin{equation}
 \begin{aligned}
 c_s&=\hat u_H^{(s)T}\hat u_F^{(s)},\\
 \zeta_s&=\sigma_{\min}[\hat u_S^{(s)},\hat u_H^{(s)},\hat u_F^{(s)}],\\
 \mathcal A_{a\to b}&=\zeta_a/\zeta_b.
 \end{aligned}\label{eq:rotation}
\end{equation}
$\hat U_s=[\hat u_S^{(s)},\hat u_H^{(s)},\hat u_F^{(s)}]$ is the normalized response matrix for the three radioactive groups. Its smallest singular value $\sigma_{\min}$ measures the weakest separation among the three unit columns; $\|\cdot\|_2$ is the Euclidean norm, and $c_s$ is the long alpha--fission response cosine. The fraction $f_F^{(s)}(t)$ uses the corresponding same stage total power or luminosity. Column normalization isolates shape from absolute photon count, so $\mathcal A$ describes the amplification of shape ambiguity. The normalized curves in Fig.~\ref{fig:master}(a--c) and their eighteen-epoch response geometry show a coherent progression: long alpha and fission responses become closely aligned during deposition and remain aligned after escape. The corresponding amplifications are $\mathcal A_{\rm raw\to dep}=3.96$ and $\mathcal A_{\rm control\ photon\to HN\ photon}=4.06$. The short $A=222$ and 224 responses are already almost parallel ($c>0.999$). Because fractional angles also include the beta-dominated denominator, their significance for attribution is resolved by profiling the bounded competing models. The two radiation group extension in Sec.~\ref{sec:twogroup} further connects this alignment to the information carried by emerging color, separating total luminosity similarity from spectral response contrast.

To isolate the deposition effect on source separation, we hold the late fission luminosity share fixed. The unit deposition control uses contemporaneous depositable charged particle and gamma energy and the same shell solver at all 27 mass, velocity and opacity templates. At $\Yatot=10^{-5},C=2\times10^{-6}$, its 18 point bounded no-$^{254}\mathrm{Cf}$ distance is 0.747, compared with 0.090 for particle dependent deposition. Rescaling the control Cf abundance to the latter's day 90 fission fraction, 0.380379, while preserving the alpha abundance, templates, errors and coefficient bounds, gives $D_G^2=0.445$ and a maximum residual of 2.94\%, versus 0.090 and 1.93\% in the particle dependent calculation. The factor 4.94 difference in squared distance survives the matching of late fission share. Deposition therefore alters the response shapes and their available compensations, in addition to its effect on source amplitude.

The matched-share comparison can also be evaluated with a causal constant stopping power treatment of alpha particles and fission fragments, motivated by the analytic energy-loss description in Ref.~\cite{Barnes2016}. For a monoenergetic species with initial energy $E_p$ and loss rate $\psi_p\rho$, an expanding uniform sphere gives
\begin{equation}
 \begin{aligned}
 q_{p,\rm dep}(t)&=\psi_p\rho(t)\int_{t_{\min}(t)}^t\frac{q_{p,\rm raw}(u)}{E_p}\,du,\\
 t_{\min}(t)&=\left[t^{-2}+\frac{2E_p}{\psi_p\rho_*}\right]^{-1/2}.
 \end{aligned}
 \label{eq:stopping}
\end{equation}
Here $\rho(t)=\rho_*t^{-3}$, with $\rho_*=3M/(4\pi v_{\rm out}^3)$ and uniform-sphere outer speed $v_{\rm out}$; $t$ and $u$ are in seconds for the quoted stopping coefficients. The lower integration limit is also clipped at the 0.01 d initialization. This propagates each actual source history rather than multiplying an exponential decay by a power-law heating efficiency. We use $E_\alpha=6$ MeV, $E_{\rm ff}=90.8$ MeV, and $\psi_\alpha=5\times10^{11}$ and $\psi_{\rm ff}=5\times10^{13}$ MeV cm$^3$ g$^{-1}$ s$^{-1}$. The sphere has the same mass and kinetic energy as each shell template: $\langle v^2\rangle=(24/7)v_{\min}^2$ and outer speed $\sqrt{5/3}\,\sqrt{\langle v^2\rangle}$. Electron and gamma deposition retain the mass- and velocity-dependent HN calculation. This comparison changes the alpha/fragment loss law and spatial averaging while keeping the emitted powers, beta treatment and gray photon solver fixed. It omits charged-particle adiabatic loss and uses representative energies; it is an explicit sensitivity model rather than an independently calibrated improvement.

\begin{figure*}[t]\centering
\includegraphics[width=0.485\textwidth]{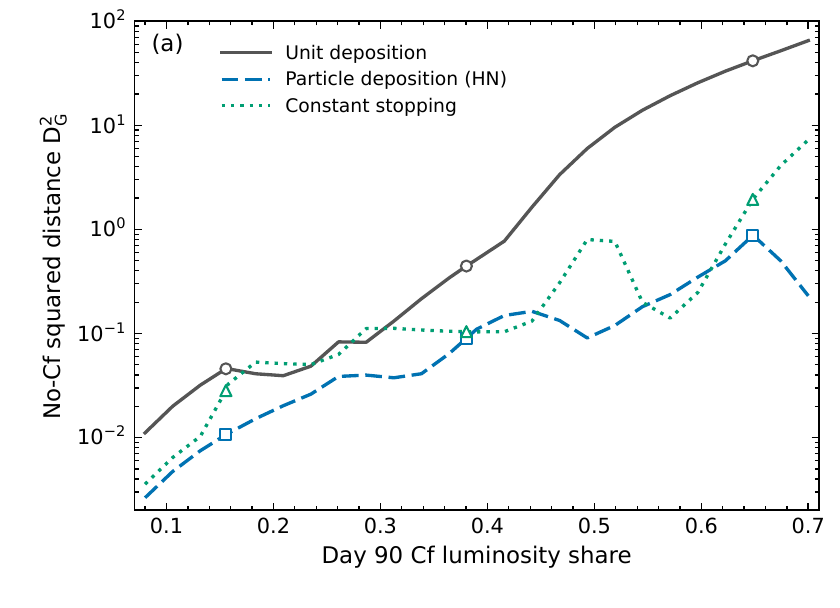}\hfill\includegraphics[width=0.485\textwidth]{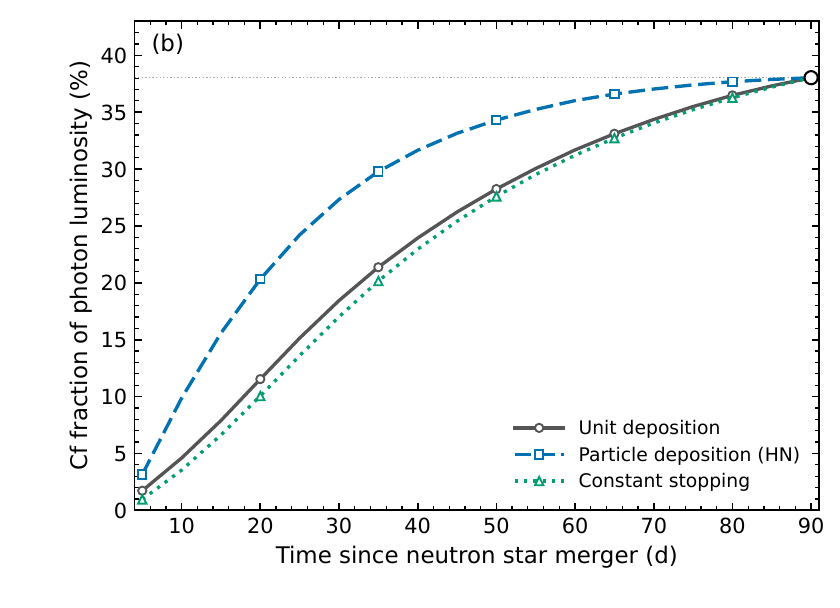}
\caption{The same late fission contribution can have different temporal signatures. (a) The profiled no-Cf distance at twenty-eight matched day 90 shares for three deposition prescriptions; open symbols mark the three reference calculations on each curve. (b) The Cf photon luminosity fraction as a function of time when all prescriptions are normalized to the same 38.04\% day 90 share. The common endpoint is marked by an open circle and a horizontal reference. The distinct earlier contributions show why matching the late share controls amplitude without equating the full radiation history. Straight segments join evaluated models; the stopping prescription is the controlled comparison specified in the text.}\label{fig:share}\end{figure*}

At the principal matched 38.0\% share, this alternative gives $D_G^2=0.1040$ and a maximum residual of 2.05\%. The preferred allocation changes, while a close no-Cf response remains available. Figure~\ref{fig:share}(a) compares twenty-eight evaluated matched shares, including three reference anchors. The injected inventory and the competing alpha budget determine the preferred compensation and the resulting deposition effect. Figure~\ref{fig:share}(b) makes the matching physically explicit: all three photon contributions meet at day 90, while their earlier luminosity fractions remain different. The comparison therefore controls a declared late source amplitude and resolves the temporal freedom retained by deposition.

\subsection{Coupled ejecta and source responses}
Ejecta density couples heating and escape through particle thermalization. At fixed abundances, $\partial_xL=(\partial_x\mathcal T)_q[q_{\rm dep}]+\mathcal T[\partial_xq_{\rm dep}]$, with $x=\ln M$ or $\ln v$ and opacity fixed where appropriate. Symmetric perturbations $\Delta\ln M=\pm0.035$ and $\Delta\ln v=\pm0.035$ estimate the log responses; velocity changes preserve the dimensionless profile. Full minus frozen responses, $\Delta_M$ and $\Delta_v$, isolate deposition coupling in Fig.~\ref{fig:coupling}(a). Recomputing all six sources gives a mass correction of 0.674 at 90 d and a velocity correction of $-1.852$ at 60 d. Their magnitude exceeds numerical variations and follows the density dependence of thermalization \cite{Barnes2016,Hotokezaka2020}, changing the available source compensation.

\subsection{Calorimetric memory}
\begin{figure*}[t]\centering
\includegraphics[width=0.485\textwidth]{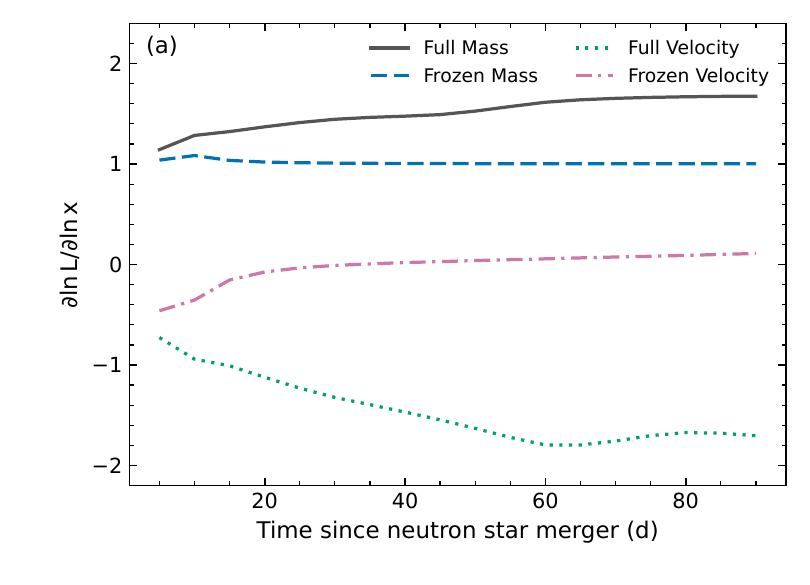}\hfill\includegraphics[width=0.485\textwidth]{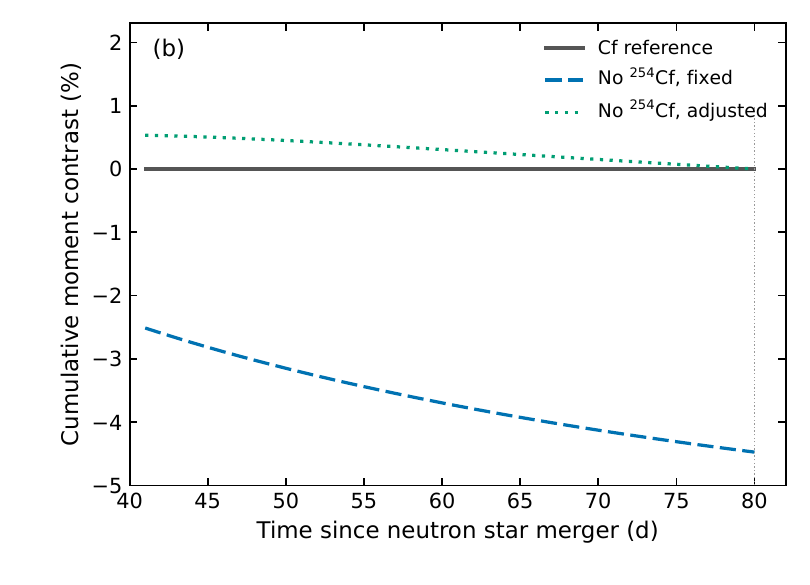}
\caption{Density coupling and radiation storage affect complementary parts of the inverse problem. (a) Full mass and velocity derivatives recompute both deposition and photon escape; frozen-deposition derivatives isolate their transport contribution. (b) Cumulative moment contrasts retain the temporal differences between a Cf reference and no-Cf alternatives, even when the adjusted alpha mixture meets the same 80 d moment anchor. The common scalar moment therefore leaves distinct light-curve evolution accessible to a time-resolved comparison.}\label{fig:coupling}\end{figure*}

Time integrated luminosity retains information about deposited heat and the release of stored radiation. Its usefulness for identifying an isotope depends on how uniquely a source inventory determines that integral. Summing Eq.~(\ref{eq:transport}) over shells, with $E=\sum_sE_s$, $Q=\sum_sM_s\qth$ and $L=\sum_sL_s$, gives the established time weighted energy identity \cite{Hotokezaka2020,KasenBarnes2019},
\begin{equation}
 \frac{d(tE)}{dt}=t[Q(t)-L(t)].\label{eq:katz}
\end{equation}
For window endpoints $0<a<b<c$, $H_{ab}=\int_a^btQdt$ and $W_{ab}=\int_a^btLdt$, define the finite window memory fraction $\mathcal B_{ab}=[bE(b)-aE(a)]/H_{ab}$. Then
\begin{equation}
 W_{ab}=H_{ab}(1-\mathcal B_{ab}),\quad
 \frac{W_{bc}}{W_{ab}}=\frac{H_{bc}}{H_{ab}}
 \frac{1-\mathcal B_{bc}}{1-\mathcal B_{ab}}.
 \label{eq:memory}
\end{equation}
A negative $\mathcal B_{ab}$ means stored radiation supplements contemporaneous heating. For the illustrative $\Yazero{}=1.35\times10^{-4}$, $C_0=2\times10^{-6}$ reference, $\mathcal B_{20,40}=-0.0370$, $\mathcal B_{40,80}=-0.0825$ and $W_{40,80}/W_{20,40}=0.81896$. Direct integration closes Eq.~(\ref{eq:katz}) within $3.9\times10^{-6}$ over the evaluated opacity changes. Doubling or halving either opacity changes the ratio by at most $0.66\%$; removing Cf reduces it to 0.78233, a 4.47\% change. Yet a no-Cf model with short alpha amplitude 0.72451 of its reference value recovers the same ratio. Figure~\ref{fig:coupling}(b) displays their common 80 d anchor and distinct earlier evolution. The moment retains calorimetric information while discarding temporal distinctions useful for attribution.

\section{Bounded source attribution}
\subsection{Local response geometry}\label{sec:local}
Local response projection identifies the part of a radioactive signature that survives adjustment of competing parameters. Write the fractional luminosity change as $\delta\ln L=S\theta+N\eta$, with three fractional source amplitude shifts in $\theta$ and six nuisance coefficients in $\eta$. The source columns $S_{kj}=L_j(t_k)/L(t_k)$ represent short alpha, long alpha and Cf amplitudes. The nuisance matrix contains flux normalization, beta tilt, and self-consistently recomputed mass, velocity and two opacity directions. These describe smooth uncertainty in the beta luminosity response, density and expansion, and inner/outer gray escape. They define an explicit sensitivity basis, which can be extended or restricted by correlated yields and external ejecta measurements. This calculation uses the illustrative $\Yazero{}=1.35\times10^{-4}$ and $C_0=2\times10^{-6}$ reference. The Jacobian has nuisance rank six for both eight and eighteen epochs, with smallest column normalized nuisance singular values 0.00472 and 0.00781. Its normalization column is one; the beta tilt column is $(L_\beta/L)\ln[t/(20\,\mathrm d)]$, and the remaining columns are log luminosity derivatives with respect to $M,v,\kappa_{\rm in},\kappa_{\rm out}$. At independent log luminosity error $\sigma$, the projected source information is\begin{equation}
 F_{\theta|\eta}=\sigma^{-2}S^T(I-NN^+)S,
 \quad \operatorname{rank}F_{\theta|\eta}\leq n-\operatorname{rank}N.
 \label{eq:fisher}
\end{equation}
Here $n$ counts epochs, $I$ is the $n\times n$ identity, and $N^+$ is the Moore--Penrose pseudoinverse; $NN^+$ projects onto the nuisance span. Six nuisance directions leave at most two source directions at eight epochs. The eight epoch reference, 5, 10, 15, 20, 25, 40, 60 and 90 d, therefore has a zero weakest source eigenvalue. Additional epochs restore rank, although the weakest projected direction remains small. The finite comparisons below determine which compensations remain accessible under positivity and abundance bounds.

Beyond the cadence dependence, self-consistent deposition materially changes this local information. For the same eighteen epoch reference, replacing frozen deposition mass and velocity columns with the full responses reduces the profiled Cf squared contrast at 10\% error from $1.73\times10^{-4}$ to $1.92\times10^{-5}$. The approximately ninefold reduction measures the additional compensation introduced by ejecta--heating coupling. 

We use $D^2=9$ as a fixed Gaussian contrast reference squared separation, corresponding to three noise standard deviations. Section~\ref{sec:two} calibrates the specified two component source deletion experiment with synthetic trials.

Local projection describes infinitesimal compensation in the unrestricted nuisance span around $\Yazero{}$. Finite deletion instead profiles positive abundances and restricted ejecta around its stated truth. The two distances describe different references and admissible displacements: local geometry identifies compensation directions, while bounded fitting checks whether they produce an accessible competing light curve.

\subsection{Finite competing inventories}\label{sec:bounded}
Substantial late Cf power can coexist with close beta and alpha alternatives. The illustrative yields $(Y_{222},Y_{223},Y_{224},Y_{225})=(4.0,2.7,4.1,2.7)\times10^{-5}$ sum to $\Yazero{}=1.35\times10^{-4}$, approximately 3.0\% by mass, and normalize the source kernels. The principal cap $\Yatot=10^{-5}$ follows the scale of the model dependent GW170817 ceiling \cite{Wu2019}; its illustrative $\Ycf=2\times10^{-6}$ injection supplies our Cf reference. The specified one and two component truths give day-90 Cf luminosity fractions of approximately 38--42\%. Other production and thermalization choices can produce fission-dominated late heating \cite{Zhu2018}. These scenarios distinguish a source's radiative contribution from the uniqueness of its attribution.

A template is a complete forward response at fixed ejecta parameters; a witness is an explicit feasible competing light curve, with all fit coefficients within their bounds. Let $B_e,S_e,H_e,F_e$ denote independently deposited and transported beta, short ($A=222,224$), long ($A=223,225$), and Cf luminosities for ejecta $e$. At fiducial $e_0$, the stipulated truth is $T=B_0+(\Yatrue{}/\Yazero{})(S_0+H_0)+(C/C_0)F_0$, $C_0=2\times10^{-6}$. The no-$^{254}\mathrm{Cf}$ alternative sets $Y_{254}=0$ and retains the beta and alpha sources. This designation concerns the modeled isotope; other fissioning nuclei would require additional response columns. The alternative is
\begin{equation}
 \begin{aligned}
 N_e(t)={}&[b+h\ln[t/(20\,\mathrm d)]]B_e(t)\\
 &+\frac{\Yacap{}a_S}{8.1\times10^{-5}}S_e(t)
 +\frac{\Yacap{}a_H}{5.4\times10^{-5}}H_e(t),
 \end{aligned}
 \label{eq:null}
\end{equation}
The denominators in Eq.~(\ref{eq:null}) are the reference grouped alpha yields; $\Yacap{}a_S$ and $\Yacap{}a_H$ are the alternative abundances. Here $b$ is the beta normalization and $h$ its additive logarithmic shape coefficient, distinct from branching probabilities. We impose $b\in[0.8,1.2]$, $|h|\leq0.2$, $a_S,a_H\geq0$ and $a_S+a_H\leq1$. Matched budgets use $\Yatrue{}=\Yacap{}=\Yatot$; varying the cap at fixed injection isolates model flexibility. The multiplier acts on the transported beta luminosity $B_e(t)$, with normalization $b$ defined at the 20 d pivot and $h$ controlling its smooth temporal variation. The coefficient bounds give pointwise factors 0.523--1.477 at 5 d and 0.499--1.501 at 90 d; the same $b,h$ link the factors across epochs.

For independent 10\% flux errors, the Asimov distance (squared residual distance for noise free synthetic data evaluated with the stipulated noise weights) is
\begin{equation}
 D_G^2(\Yatrue{},\Yacap{},C)=\min_{e\in G,b,h,a_S,a_H}
 \sum_i\left[\frac{T_i-N_{e,i}}{0.1T_i}\right]^2.
 \label{eq:dg}
\end{equation}
Here $T_i=T(t_i)$ is the synthetic reference, $N_{e,i}=N_e(t_i)$ the competing luminosity, and $G$ the finite template set. At each template, the coefficient fit is a convex quadratic over the beta bounds and alpha simplex. Deposition is recomputed for masses $0.015,0.023,0.035\Msun$, minimum speeds $0.08,0.10,0.12c$, and opacity multipliers $0.5,1,2$, giving 27 templates. Enlarging the family preserves each feasible alternative, so $D_{\rm full}^2\leq D_G^2$. A low distance thus supplies an explicit competing explanation. Equation~(\ref{eq:powerbound}) interprets separation statistically; the $D^2=9$ contour remains a fixed-pair reference, and Sec.~\ref{sec:two} separately calibrates the two component deletion statistic.

\begin{figure*}[t]\centering
\includegraphics[width=0.485\textwidth]{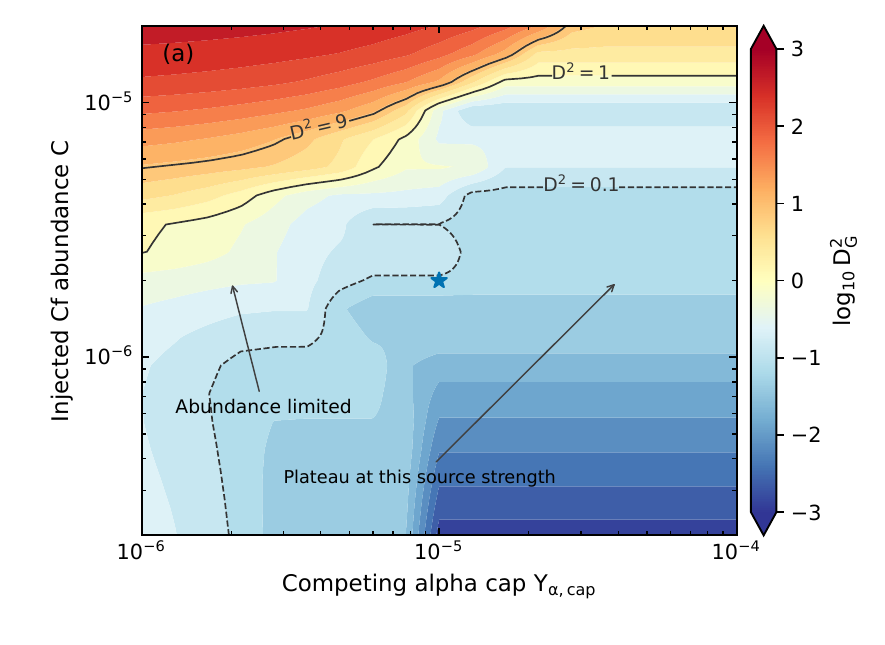}\hfill\includegraphics[width=0.485\textwidth]{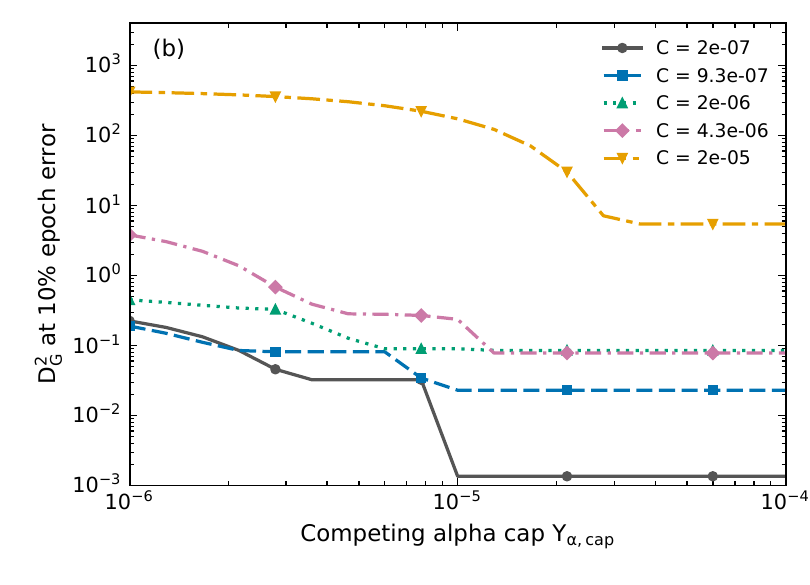}
\caption{Source strength and competing yield freedom control different regimes of attribution. (a) At fixed injected alpha abundance $10^{-5}$, the separation map distinguishes an abundance-limited alternative from the plateau reached once its preferred alpha budget is available. The star marks the matched-budget reference; the arrows identify the regimes at the indicated source strength. (b) Evaluated cap slices resolve that transition directly. The 361 model evaluations are distinct from the bilinear interpolation in logarithmic coordinates. Contour labels give squared distances, not detection thresholds.}\label{fig:phase}\end{figure*}

Ejecta freedom opens close competing responses (Fig.~\ref{fig:phase}). At $\Yatot=10^{-5},C=2\times10^{-6}$, eighteen epochs at 5 d spacing give fixed ejecta distance $D_{\{e_0\}}^2=0.954$ and profiled distance $D_G^2=0.090$. The preferred $(M,v_{\min},\kappa/\kappa_0)=(0.035\Msun,0.12c,2)$ has $(b,h,a_S,a_H)=(0.8766,0.1397,0,0.5574)$ and maximum residual 1.93\%, despite a 38.0\% day-90 Cf share. At $\Yatot=10^{-6}$, profiling reduces the fixed ejecta distance 32.57 to 0.517 with tilt and 6.63 without. The fixed ejecta $D^2=9$ contour moves from $C=1.60\times10^{-6}$ at $\Yatot=10^{-6}$ to $1.12\times10^{-5}$ at $\Yatot=1.35\times10^{-4}$. Source budget and ejecta variation jointly control the compensation shown in Fig.~\ref{fig:master}(d).

The $19\times19$ map fixes $\Yatrue{}=10^{-5}$ and varies $C$ and $\Yacap{}$ using the same epochs, 27 templates and beta bounds. Increasing the cap enlarges a nested feasible set, so the minimum distance is non-increasing. All nineteen Cf rows obey this ordering. The 361 recomputed fits resolve an abundance limited regime and a plateau where the preferred allocation lies within the budget; contours interpolate them in logarithmic coordinates. Dual certificates independently verify the principal grouped and four-seed minima.

The plateau further distinguishes total abundance freedom from isotope allocation freedom. At $C=2\times10^{-6}$, the grouped distance is 0.08997 for $\Yacap{}=10^{-5}$ and reaches 0.08526 once the budget accommodates the preferred total alpha yield $1.275\times10^{-5}$. Independent chain allocation yields the distinct four seed solution below, with distance 0.02942 under the original $10^{-5}$ cap. A tighter total budget and stronger correlations among chain yields therefore constrain different parts of the inverse problem. The map quantifies the influence of the total abundance budget and provides a baseline for incorporating correlations among nuclear yields \cite{Wu2019,Zhu2021,Pognan2025}.

Independent alpha allocation gives a closer no-$^{254}\mathrm{Cf}$ light curve at the same total cap. The grouped family ties $A=222$ to 224 and 223 to 225; the enlarged family allows four $Y_j\geq0$ with $\sum_jY_j\leq \Yacap{}$, $0\leq Y_F\leq C$, and the same beta bounds. Source deletion sets the selected yield to zero while retaining the injected mixture and 27 templates. Deleting Cf gives $D_G^2=0.0294$ and maximum residual 1.11\% at a 38.0\% day-90 Cf share. Even at $h=0$, grouped and independent allocations give $D_G^2=0.359$ and 0.113, with 14.79\% and 9.53\% Gaussian power ceilings at 5\% false positives and 10\% epoch errors (Fig.~\ref{fig:master}(d)). Thus alpha allocation and ejecta already supply compensation before beta shape freedom is added. Nuclear production correlates these yields through electron fraction, entropy, expansion, captures, beta feeding, odd--even structure and fission recycling \cite{Wu2019,Zhu2021,Pognan2025}. For any correlated subfamily at the same ejecta, $D_{\rm independent,min}^2\leq D_{\rm correlated,min}^2$; the explicit independent allocation characterizes the enlarged family, and its nuclear realization requires joint yields.

The grouped and four-seed fits have effective beta factors 0.683 and 0.835 at 5 d, rising to 1.087 and 1.006 at 90 d. Their tilts, 0.140 and 0.059, lie inside $|h|\leq0.2$, while at least one alpha yield reaches zero. These phenomenological coefficient bounds define a time dependent flux range; calibration as a nuclear uncertainty distribution uses joint heating histories and yields.

\subsection{Certified source separation}\label{sec:finite}
A feasible alternative constrains source attribution before any particular recovery statistic is chosen. This connection becomes exact for Gaussian flux errors with one known covariance $\Sigma$ shared by the compared means. The covariance is held fixed at the stated synthetic reference; assigning different error matrices to each hypothesis would define a different experiment. For an ejecta template $e$, let $\mathsf K_e$ contain its beta, beta-shape and alpha response columns and let $x$ contain their coefficients. The box bounds and positive abundance simplex define a compact convex set $\mathcal P$. Writing $W^TW=\Sigma^{-1}$, $z=WT$ and $\mathsf M_e=W\mathsf K_e$ gives
\begin{equation}
 d_e^2=\min_{x\in\mathcal P}\|z-\mathsf M_ex\|_2^2,
 \qquad d_G^2=\min_{e\in G}d_e^2.
 \label{eq:finiteprojection}
\end{equation}
This is the general-covariance form of Eq.~(\ref{eq:dg}). The union over ejecta need not be convex; each individual coefficient problem is convex and can be checked separately \cite{Boyd2004}.

An explicit dual certificate verifies its minimum. Define the support function $s_{\mathcal P}(p)=\max_{x\in\mathcal P}p^Tx$. The completed square $\|r\|^2=\|r-v\|^2+2v^Tr-\|v\|^2$, with $r=z-\mathsf M_ex$, supplies a lower bound for any data-space vector $v$:
\begin{equation}
 d_e^2\geq 2v^Tz-\|v\|_2^2
       -2s_{\mathcal P}(\mathsf M_e^Tv).
 \label{eq:dualseparation}
\end{equation}
For independent box coefficients indexed by $\mathcal J$, with lower and upper bounds $\ell_j,u_j$, and alpha coefficients indexed by $\mathcal I_\alpha$ in $x_j\geq0$, $\sum_{j\in\mathcal I_\alpha}x_j\leq1$, the support is available in closed form:
\begin{equation}
 s_{\mathcal P}(p)=\sum_{j\in\mathcal J}\max(\ell_jp_j,u_jp_j)
          +\max\!\left(0,\max_{j\in\mathcal I_\alpha}p_j\right).
 \label{eq:support}
\end{equation}
At the projection, $v=z-\mathsf M_e\hat x$ saturates Eq.~(\ref{eq:dualseparation}): $v^T\mathsf M_e(x-\hat x)\leq0$ for feasible $x$ implies $s_{\mathcal P}(\mathsf M_e^Tv)=v^T\mathsf M_e\hat x$. Minimizing the lower bounds across all templates checks the full finite library. For the grouped and four-seed distances 0.08996 and 0.02942, primal and dual values agree within $10^{-10}$ and $10^{-12}$, respectively. Each feasible solution also supplies an upper bound on separation in a larger family.

\subsection{Detection power of admissible source alternatives}
The statistical implication follows from the Gaussian likelihood ratio \cite{Neyman1933}. Consider a Cf truth $T$ and one feasible no-Cf mean $N$. Set $\delta=T-N$ and $d^2=\delta^T\Sigma^{-1}\delta$. The score $Z=\delta^T\Sigma^{-1}(y-N)/d$ has distributions $\mathcal N(0,1)$ under $N$ and $\mathcal N(d,1)$ under $T$. Its optimal level-$\alpha_{\rm FP}$ threshold is $z_{1-\alpha_{\rm FP}}$, where $z_p=\Phi^{-1}(p)$ and $\Phi$ is the standard normal cumulative distribution. Therefore every decision rule whose false-positive probability is at most $\alpha_{\rm FP}$ for every mean in the null family obeys
\begin{equation}
 \Pr_T(\hbox{identify Cf})\leq
 \Phi\!\left(d_G-z_{1-\alpha_{\rm FP}}\right).
 \label{eq:powerbound}
\end{equation}
Uniform false-positive control holds at the closest null in particular, giving the composite-family ceiling. Any feasible witness also gives a valid, potentially weaker bound through its own distance.

At 5\% false positives, the grouped and four-seed distances give power ceilings of 8.93\% and 7.03\%; the two component distance 0.753 gives 21.85\%. Simulations with 200,000 realizations per pair reproduce the analytic envelopes at 1\%, 5\% and 10\% false-positive levels. These limits concern the specified means, covariance and inventories. Enlarging the alternative family at fixed reference preserves them through a non-increasing minimum distance; changing the reference physics defines a new comparison.

An intended power $\pi_*$ consequently requires
\begin{equation}
 d_G^2\geq\left[z_{1-\alpha_{\rm FP}}+z_{\pi_*}\right]^2,
 \label{eq:necessarypower}
\end{equation}
when the bracket is positive. This necessary condition sets the precision needed to permit the target power. For unchanged means, source family and cadence, uniform independent errors give $d_G^2(\sigma)=d_G^2(\sigma_0)(\sigma_0/\sigma)^2$ and power ceiling $\Phi[\sqrt{d_G^2(\sigma_0)}\,\sigma_0/\sigma-z_{1-\alpha_{\rm FP}}]$. The inverse-square law applies to squared distance; power follows the Gaussian expression. Unequal errors or changed covariance require a new weighted comparison. The four-seed witness requires per-epoch error at most 0.69\% to permit 80\% power at 5\% false positives under the stated model.

\subsection{Collective limits from competing source inventories}\label{sec:collective}
Several admissible inventories can jointly tighten the attribution limit. We derive a collective likelihood bound that enforces false-positive control at every retained competing inventory.

Let $P_T=\mathcal N(T,\Sigma)$ and let $P_j=\mathcal N(N_j,\Sigma)$ describe $m$ feasible no-Cf inventories with the same fixed covariance. For a decision function $0\leq\varphi(y)\leq1$, impose $\mathbb E_j\varphi\leq\alpha_{\rm FP}$ for each $j$. Write $\ell_j=dP_j/dP_T$ and $r_\theta=\sum_j\theta_j\ell_j$, with nonnegative $\theta_j$. The collective inequality can be established by separating the likelihood weighted contribution from its remainder. The change of measure and the bound $\varphi(1-r_\theta)\leq(1-r_\theta)_+$ give
\begin{equation}
\begin{aligned}
 \mathbb E_T\varphi
 &=\mathbb E_T[\varphi r_\theta]
    +\mathbb E_T[\varphi(1-r_\theta)]\\
 &=\sum_j\theta_j\mathbb E_j\varphi
    +\mathbb E_T[\varphi(1-r_\theta)]\\
 &\leq\alpha_{\rm FP}\sum_j\theta_j
    +\mathbb E_T[(1-r_\theta)_+],
\end{aligned}\label{eq:collective}
\end{equation}
where $(a)_+=\max(a,0)$.
Every feasible weight vector gives a valid upper bound; minimizing strengthens it. One alternative with an optimized scalar weight recovers Eq.~(\ref{eq:powerbound}) \cite{Neyman1933}.

The Gaussian model makes the bound directly computable. For $Z=\Sigma^{-1/2}(y-T)\sim\mathcal N(0,I)$ and $u_j=\Sigma^{-1/2}(T-N_j)$,
\begin{equation}
 \ell_j(Z)=\exp[-u_j^TZ-\tfrac12\|u_j\|^2].
 \label{eq:gaussianmixture}
\end{equation}
Writing $\eta=\sum_j\theta_j$ and $\omega_j=\theta_j/\eta$ for $\eta>0$ defines a statistical mixture $\sum_j\omega_jP_j$ for bounding the decision rule. Its members remain distinct physical source configurations.

A closed-form collective bound follows by retaining the geometry of the same Gaussian source responses. For normalized weights $\omega_j\geq0$, $\sum_j\omega_j=1$, define the mean response displacement $\bar u_\omega=\sum_j\omega_ju_j$, its length $d_\omega=\|\bar u_\omega\|$, and the response dispersion
\begin{equation}
 V_\omega=\sum_j\omega_j\|u_j\|^2-d_\omega^2,
 \qquad\gamma_\omega=e^{-V_\omega/2}.
 \label{eq:dispersion}
\end{equation}
The weighted arithmetic--geometric mean inequality yields
\begin{equation}
\begin{aligned}
 \sum_j\omega_j\ell_j(Z)
 &\geq \prod_j\ell_j(Z)^{\omega_j}\\
 &=\gamma_\omega\exp[-\bar u_\omega^TZ-\tfrac12d_\omega^2].
\end{aligned}\label{eq:geometricminorant}
\end{equation}
The exponential on the right is the likelihood ratio of a Gaussian reference with mean $\bar N_\omega=\sum_j\omega_jN_j$ and covariance $\Sigma$. For any uniformly valid decision, nonnegativity of $\varphi$ then gives $\gamma_\omega\mathbb E_{\bar N_\omega}\varphi\leq\sum_j\omega_j\mathbb E_j\varphi\leq\alpha_{\rm FP}$. Applying the Gaussian likelihood ratio envelope at level $\alpha_{\rm FP}/\gamma_\omega$ therefore gives
\begin{equation}
 \Pr_T(\hbox{identify Cf})\leq
 \Phi\!\left(d_\omega-z_{1-\alpha_{\rm FP}/\gamma_\omega}\right),
 \label{eq:analyticcollective}
\end{equation}
for $\alpha_{\rm FP}<\gamma_\omega$; otherwise the bound is one. Concentrating weight on one inventory sets $V_\omega=0$ and recovers Eq.~(\ref{eq:powerbound}). The mathematical Gaussian reference accounts for both the mean displacement and dispersion in the covariance metric; it need not be a physical ejecta template.

For the grouped and four seed comparisons, we retain respectively ten and eleven feasible witnesses with squared distance at most four. A pilot sample of 12,000 independent Gaussian draws makes the positive-part optimization a finite linear program. Seven nonnegative weight vectors are selected for each family and then held fixed. An independent sample of $n_{\rm MC}=2\times10^6$ draws evaluates each positive-part expectation. Since its integrand is in $[0,1]$, adding
\begin{equation}
 \epsilon_{\rm MC}=\sqrt{\frac{\ln(n_{\rm int}/\delta_{\rm MC})}{2n_{\rm MC}}}
 \label{eq:mixtureMC}
\end{equation}
to each sample mean gives simultaneous upper coverage at least $1-\delta_{\rm MC}$ for $n_{\rm int}$ evaluated integrals by the bounded-variable inequality and a union bound \cite{Hoeffding1963}. We use $n_{\rm int}=14$ and $\delta_{\rm MC}=0.001$, giving $\epsilon_{\rm MC}=0.001545$. Weight selection and validation use independent draws. The coverage concerns numerical integration of the bound, separately from the false-positive level of an observational decision.

At 5\% false positives, Eq.~(\ref{eq:analyticcollective}) gives 7.33\% and 6.20\% ceilings for grouped and independent alpha seeds, compared with single-witness limits 8.93\% and 7.03\%. Numerical evaluation including the allowance gives 7.16\% and 6.19\%. One-dimensional quadrature checks the closed form, and 100,000 independent score vectors check Eq.~(\ref{eq:geometricminorant}) pointwise for each family. Figure~\ref{fig:finite}(a) shows how the allowed collection tightens the attribution limit.

\subsection{Independent heating and yield constraints}
\begin{figure*}[t]\centering
\includegraphics[width=0.485\textwidth]{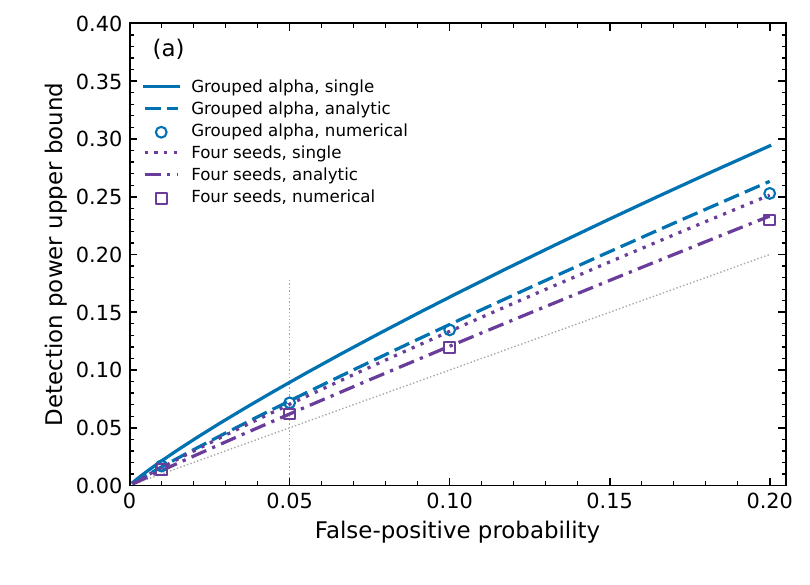}\hfill\includegraphics[width=0.485\textwidth]{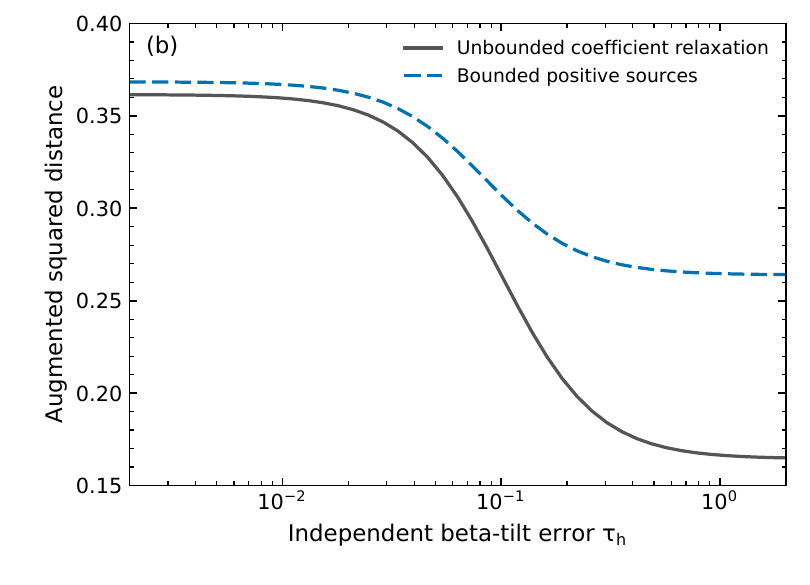}
\caption{Several admissible inventories can tighten the limit on isotope identification. (a) Line patterns compare single-witness Gaussian ceilings with the closed-form collective ceilings for grouped and independently allocated alpha seeds. Open circles and squares show the corresponding numerical collective bounds, including simultaneous 99.9\% integration allowances. All curves and symbols are upper bounds rather than achieved recovery fractions. The dotted diagonal gives the no-separation reference. (b) Bounded fits and coefficient relaxations use the same truth, covariance, ejecta templates and independent Gaussian beta information. Their ordered curves isolate the role of source constraints without changing the physical reference.}\label{fig:finite}\end{figure*}

The same coefficient family also connects finite fits to their linear relaxation. Independent Gaussian information with mean $\mu$ and precision factor $R$ replaces $\mathsf M_e,z$ by
\begin{equation}
 \widetilde{\mathsf M}_e=\begin{pmatrix}\mathsf M_e\\R\end{pmatrix},
 \qquad\tilde z=\begin{pmatrix}z\\R\mu\end{pmatrix}.
 \label{eq:augmentedfamily}
\end{equation}
The bounded fit minimizes the augmented residual over $\mathcal P$; relaxing to $x\in\mathbb R^{n_x}$, $n_x=\dim x$, can only lower it. On an active face, $x=x_0+Vu$ reduces the stationary fit to projection onto $\widetilde{\mathsf M}_eV$ while feasibility holds. Figure~\ref{fig:finite}(b) compares both descriptions at the same truth, covariance and 27 templates, with independent beta-normalization error 0.1 and varied tilt error $\tau_h$. This sensitivity constraint isolates coefficient bounds at fixed reference; its physical nuclear calibration requires a joint ensemble.

An equally explicit source restriction fixes the short alpha fraction near the reference value 0.6. Allowing $a_S/(a_S+a_H)$ in $[\max(0,0.6-\epsilon),\min(1,0.6+\epsilon)]$ is a nested convex family, implemented as positive mixtures of its endpoint ratios; the zero-alpha allocation is included. Fixing the ratio gives $d_G^2=0.37677$, compared with 0.08996 when both groups are free. Even the fixed ratio admits at most 15.13\% power at the same false-positive level while the stipulated beta/ejecta freedom remains. This controlled source-allocation restriction isolates yield-ratio information independently of a network covariance. Network-conditioned patterns can replace it once the joint isotope yields and their production histories are available \cite{Pognan2025,Zhu2021}.

\section{Heating freedom and source recovery}\label{sec:two}
\subsection{Two component source models}
Spatially distinct ejecta give the radioactive sources different deposition and escape histories. We model this effect with two components that assign the heating channels to different physical conditions. A fast lower opacity component carries beta heating, while a slower higher opacity component carries beta, alpha and possible Cf sources. The synthetic reference uses blue $(M,v_{\min},k)=(0.015\Msun,0.12c,0.5)$ and red $(0.035\Msun,0.08c,2)$, where $k$ scales both reference gray opacities. Each source is deposited and transported in its own ejecta. The competing library combines eight blue choices $(M/\Msun\in\{0.015,0.023\},v/c\in\{0.10,0.12\},k\in\{0.5,1\})$ with eight red choices $(M/\Msun\in\{0.023,0.035\},v/c\in\{0.08,0.10\},k\in\{1,2\})$. At each of 64 pairs we profile separate beta normalizations in $[0.8,1.2]$, a common tilt $|h|\leq0.2$, and the same alpha simplex. This prescribed source assignment resolves the interaction of component structure and heating freedom; network conditioned compositions and angle dependent transfer can subsequently define a more complete family.

Smooth beta freedom controls the two component separation. At $(\Yatot,C)=(10^{-5},2\times10^{-6})$, the logarithmic fit gives $D_G^2=0.753$, compared with 16.89 at fixed beta shape (Table~\ref{tab:shape}), despite a 42.4\% day-90 Cf share. The fit reaches $h=0.2$ and $b_{\rm red}=1.2$. Blue/red beta factors are 0.585/0.923 at 5 d and 1.163/1.501 at 90 d. Their amplitude range is smaller than the approximately one-order spread in absolute beta heating across the cited low electron fraction calculations, rising to about two orders near 100 d \cite{Zhu2021}. This provides amplitude context; joint heating histories and isotope yields are needed to calibrate the correlated temporal compensation. The coefficient bounds remain a phenomenological sensitivity family. Refining the five blue/red beta, red long alpha and red Cf kernels changes luminosities by at most $0.0282\%$ at four 5--90 d anchors. The finite library establishes the stated comparisons; continuous ejecta separation requires extending it.

\subsection{Dependence on beta heating shape}
Close alternatives persist under other smooth beta shapes. At $\Yatot=10^{-5},C=2\times10^{-6}$, let $z=\ln[t/(20\,\mathrm d)]$ and $\phi_q=c_q(e^{qz}-1)/q$, $q=\pm0.25$. Normalize $c_q$ so $\max_i|\phi_q(t_i)|=\max_i|z_i|$ over the eighteen epochs, retaining the same coefficient bounds and cap. These bent bases give $D_G^2=2.045$ and 0.616. Multiplying both beta components by $[t/(20\,\mathrm d)]^h$, $|h|\leq0.2$, instead gives $D^2=0.222$; its exponent bound permits different flux excursions (Table~\ref{tab:shape}). Independent enumeration of active constraint faces, defined by the saturated coefficient bounds, verifies the winning fits within $2\times10^{-10}$ in $D^2$. A power-law change in the transported beta response gives the log direction to first order, while inventory and deposition changes motivate curvature \cite{Zhu2021,KasenBarnes2019}. Joint nuclear yields and heating histories determine which amplitudes and shapes occur physically \cite{Wu2019,Bulla2023}.

\begin{table}[htbp]\centering
\caption{Heating freedom controls the competing radiation response. All rows use the same two component truth, eighteen epochs, 10\% flux errors and 64 ejecta pairs. The beta-only row deletes alpha sources as well as Cf; the other rows retain the capped alpha inventory and delete Cf. Bounds and normalizations are specified in the text.}\label{tab:shape}
\small\setlength{\tabcolsep}{7pt}\begin{tabular}{lrr}\toprule
\colhead{Allowed heating\\response} & \colhead{Squared\\distance} & \colhead{Largest\\residual}\\\midrule
Beta only &37.799&18.02\%\\
Beta + alpha, fixed shape &16.892&21.29\%\\
Additive logarithm &0.753&3.66\%\\
Bent additive, $q=-0.25$ &2.045&5.50\%\\
Bent additive, $q=+0.25$ &0.616&3.34\%\\
Multiplicative power law &0.222&2.54\%\\\bottomrule
\end{tabular}\end{table}

The same response compensation also operates in the reverse direction. Starting from a no-$^{254}\mathrm{Cf}$ reference, we require a positive Cf yield in the competing family. This isolates physical closeness separately from noise induced likelihood improvements: an unrestricted present isotope family includes zero Cf and recovers the noiseless reference. At $\Yatot=10^{-5}$, imposing $C\geq2\times10^{-6}$ gives $D_G^2=3.966$ across the 64 pairs, while $C\geq6\times10^{-6}$ gives 9.947. These distances quantify compatibility of prescribed source inventories within the evaluated family; the sampling calculation below supplies the statistical comparison.

\begin{figure*}[t]\centering
\includegraphics[width=0.485\textwidth]{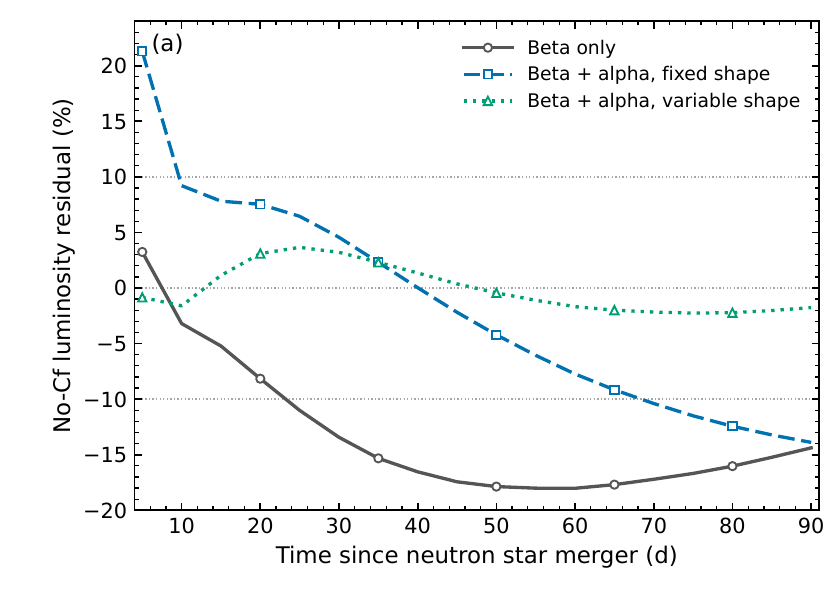}\hfill\includegraphics[width=0.485\textwidth]{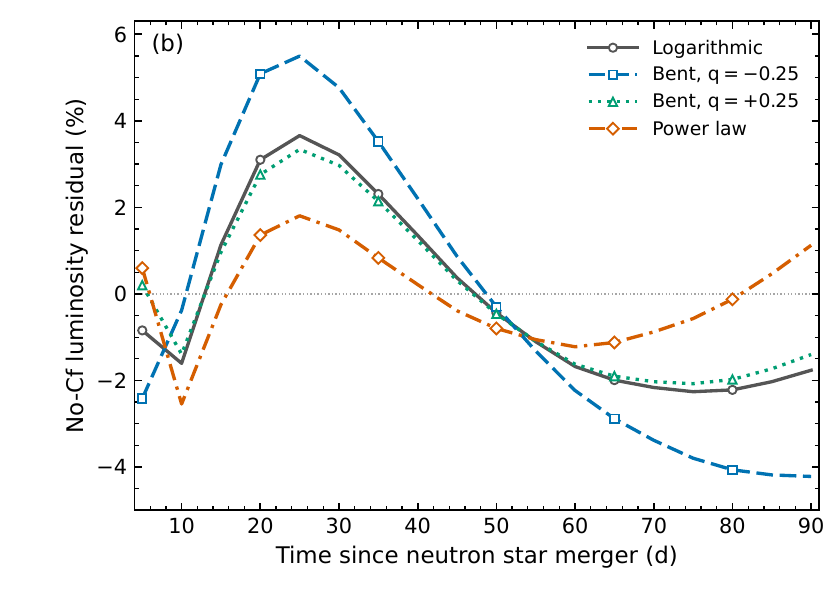}
\caption{Smooth beta-response changes can compensate a fission contribution. (a) Adding alpha responses and then allowing a smooth beta correction successively improves the no-Cf representation of the same two component truth. The horizontal $\pm10\%$ references give the assumed epoch errors. (b) An expanded residual scale resolves the distinct curvature retained by four beta prescriptions. Colors, line patterns and point symbols distinguish the evaluated alternatives; straight segments connect the eighteen calculated epochs. The different vertical scales expose both the large fixed-shape mismatch and the smaller shape-dependent residuals.}\label{fig:betafreedom}\end{figure*}

\subsection{Calibrated recovery and source deletion controls}
Statistical recovery depends on which source configurations define the null calibration. We therefore compare two explicit no-$^{254}$Cf references: the fiducial zero-Cf injection and the closest profiled alternative to the $\Yatot=10^{-5},C=2\times10^{-6}$ truth. Their source allocations and ejecta settings differ, despite their common isotope deletion. Every ensemble uses the same fixed covariance, $\sigma_i=0.1T_i$, where $T_i$ is that Cf-bearing reference; this keeps the likelihood weights identical between the geometric and sampling comparisons. We profile $C=0$ and $0\leq C\leq2\times10^{-5}$ over the same 64 template pairs and coefficient bounds, and use $\Delta\chi^2=\chi^2_{C=0}-\chi^2_{C\geq0}$.

Separate 800-trial ensembles give null 95th percentiles 0.3383 and 3.3053. Their maximum, 3.3053, sets the two-reference threshold. Independent 800-trial holdouts give exceedances 0.0\%, 5.75\% and 23.0\% at the fiducial null, closest null and Cf truth (Fig.~\ref{fig:sampling}(a)). At the measured closest-null rate 5.75\%, Eq.~(\ref{eq:powerbound}) permits 23.94\% fixed-pair power, consistent with 23.0\% recovery. The result reconciles close mean responses with the apparent recovery obtained against a selected null. Calibration at two references does not ensure uniform control over the composite null; threshold and holdout fractions retain sampling uncertainty. Standard one-parameter asymptotic thresholds require regularity that can fail for bounded coefficients and discrete templates \cite{Cowan2011}.

Deleting alpha as well as Cf and fitting the same beta terms gives $D^2=37.799$, versus 0.753 with capped alpha (Fig.~\ref{fig:betafreedom}(a)). Thus the close alternative uses both alpha responses and beta freedom. Conversely, deleting the weak $A=222$ contribution in the one component comparison gives $D^2=1.4947\times10^{-5}$ at a 0.278\% day-90 luminosity share, distinguishing weak-source redundancy from compensation of substantial Cf power.

\subsection{Cadence and correlated flux errors}
Temporal coverage and shared calibration uncertainty act on different response directions. For a predeclared common fractional flux error $\tau_c$, we use the fixed reference covariance
\begin{equation}
 \Sigma=D+\tau_c^2TT^T,\qquad D_{ii}=(0.1T_i)^2.
 \label{eq:commoncov}
\end{equation}
This is a linearized common-offset model around the stipulated truth, with the same covariance for every candidate. For residual $r=T-N$, its quadratic form is
\begin{equation}
 r^T\Sigma^{-1}r=r^TD^{-1}r-
 \frac{\tau_c^2(r^TD^{-1}T)^2}{1+\tau_c^2T^TD^{-1}T}.
 \label{eq:commonquadratic}
\end{equation}
Direct whitening agrees with this expression within $3\times10^{-14}$. It removes sensitivity along the common flux direction; it does not remove an arbitrary discrepancy at each epoch. Increasing a covariance by a positive semidefinite term cannot increase the minimized separation. In this example, a 20\% common uncertainty changes the grouped eighteen-epoch distance only from 0.089962 to 0.089640. Much of that normalization freedom is already accommodated by the fitted source coefficients, leaving the shape mismatch as the relevant residual.

A smooth calibration drift probes temporal structure beyond the common flux direction. We model it with the fixed covariance
\[
\begin{aligned}
 \Sigma_{ij}&=(0.1T_i)^2\delta_{ij}+\tau_{\rm sm}^2 T_iT_jK_{ij},\\
 K_{ij}&=\exp\!\left[-\frac{\ln^2(t_i/t_j)}{2\ell^2}\right],
\end{aligned}
\]
where $\delta_{ij}$ is the Kronecker delta, $\tau_{\rm sm}$ is the added fractional uncertainty and $\ell$ is its correlation length in logarithmic time. For $\tau_{\rm sm}=0.1$ and $\ell=0.7$, the grouped eighteen-epoch distance falls to 0.05684, compared with 0.08988 for a 10\% common offset and 0.08996 for independent errors alone. This distance gives a 7.98\% upper bound on detection power at a 5\% false-positive level under the stated Gaussian experiment. The temporal correlation therefore reduces sensitivity to part of the remaining response curvature. The same covariance is used for every candidate mean, so the likelihood bounds retain their stated interpretation.

Direct forward solutions at five illustrative model epochs, 15, 29, 45, 61 and 90 d, give $d_G^2=0.0016034$ and a 5.43\% power ceiling at independent 10\% errors. The nested three-epoch design, 15, 29 and 61 d, and two-epoch design, 29 and 61 d, admit exact bounded matches; adding 5\% or 10\% shared errors preserves these findings. With four source coefficients, the three-epoch response has rank three and exact feasible matches in 5 of 27 templates. Five epochs have rank four and no exact match. Setting $h=0$ and fixing the alpha ratio to $0.6:0.4$ raises the three- and five-epoch distances to 0.1126 and 0.4990, with 9.52\% and 17.40\% ceilings at 5\% false positives. These synthetic schedules distinguish sampling limitations from the gain supplied by source restrictions.

\begin{figure*}[t]\centering
\includegraphics[width=0.485\textwidth]{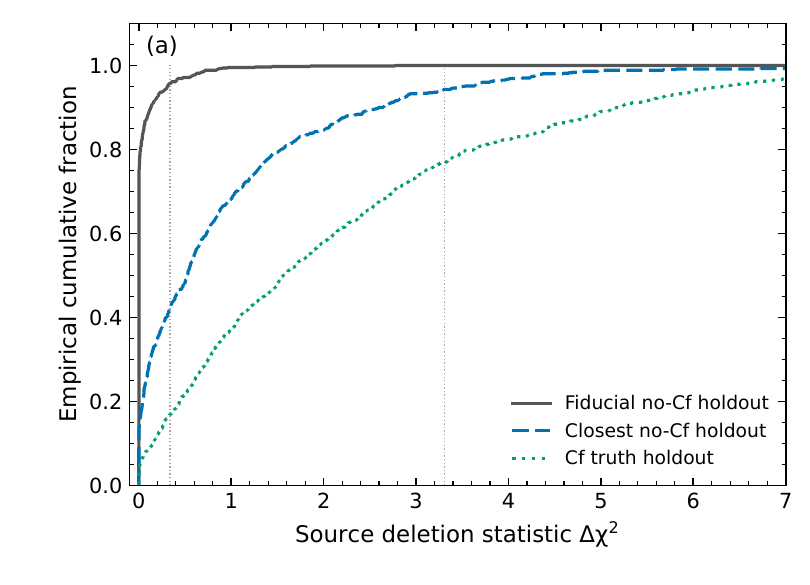}\hfill\includegraphics[width=0.485\textwidth]{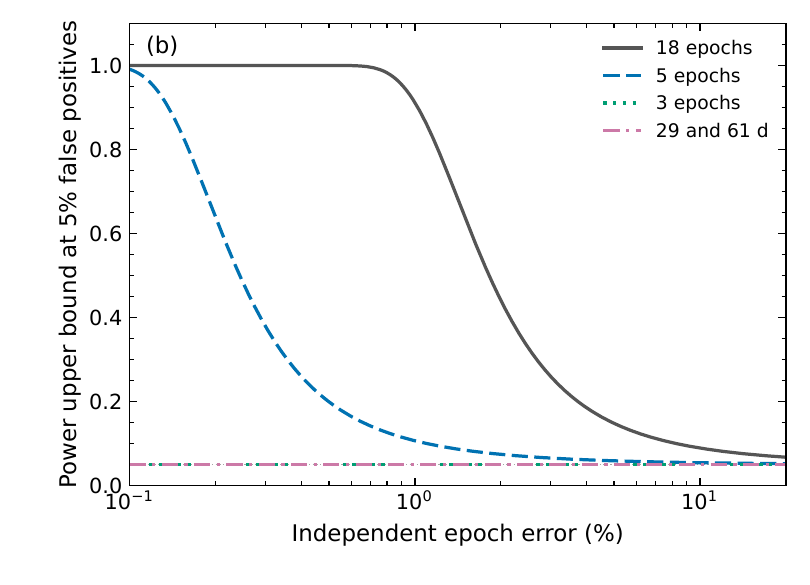}
\caption{Recovery depends on the null inventory and on the temporal information retained. (a) Independent holdout cumulative distributions from the fixed-covariance two component experiment, with the two calibrated threshold anchors. (b) Gaussian power ceilings as the independent epoch error changes for each declared schedule, with source models and bounds fixed. Three and two epochs admit numerical equality of the competing means; the corresponding curves remain at the false-positive anchor. Curves for the eighteen- and five-epoch designs follow $d_G^2\propto\sigma^{-2}$ exactly within their fixed designs.}\label{fig:sampling}\end{figure*}

An additional epoch tests the part of the source mismatch not predicted by the retained epochs and their correlated errors. For nested schedules, deleting data with its induced covariance cannot increase the separation: applying the Schur-complement identity to each fixed residual adds a nonnegative conditional square when an epoch is restored, and minimizing preserves the inequality. The two, three and five epoch schedules here are nested; the original eighteen-epoch cadence is a separate design. Figure~\ref{fig:sampling}(b) displays the exact uniform-error scaling for each declared design. The sampling and admissible response family determine which source information improved precision can recover.

\section{Independent radiation information}
\subsection{Complementary response constraints}
Independent observations improve source identification when they constrain the response that compensates the isotope of interest. This principle follows directly from local projection and provides a quantitative way to select complementary measurements. Whitening divides each response row by its independent standard error or applies the inverse square root of the joint covariance. Let $s_F$ be the whitened fission response and $X=[s_S,s_H,N_w]$, where $s_S,s_H$ are the whitened competing source responses and $N_w$ is the nuisance matrix. The information remaining after compensation is\begin{equation}
 I_{F|X}=\min_x\|s_F-Xx\|_2^2
       =\|(I-XX^+)s_F\|_2^2.\label{eq:target}
\end{equation}
An independently constrained ejecta parameter removes part of the bolometric compensation. For $X$ of full column rank, set $H_X=X^TX$ and $\hat x=H_X^{-1}X^Ts_F$. Appending one independent whitened observable with fission response $q$ and competing response row $r^T$ gives a mismatch $a=q-r^T\hat x$. Writing $x=\hat x+u$ reduces the added cost to $\min_u\{u^TH_Xu+(a-r^Tu)^2\}$, and its stationary solution yields
\begin{equation}
 I_{F|X}^{\rm new}-I_{F|X}
 =\frac{(q-r^T\hat x)^2}{1+r^TH_X^{-1}r}.
 \label{eq:gain}
\end{equation}
Here $x$ contains the compensating source and nuisance coefficients, and $e_v$ selects the $\ln v$ coefficient. An independent measurement of $\ln v$ with standard error $\tau_v$ has $q=0$ and $r=e_v/\tau_v$, giving a gain $\hat x_v^2/[\tau_v^2+(H_X^{-1})_{vv}]$. The gain is positive precisely when velocity participates in the best source compensation.

At the eighteen epoch reference of Sec.~\ref{sec:local}, with independent 10\% log flux errors, $\tau_v=0.05$ raises $I_{F|X}$ from $1.92\times10^{-5}$ to $9.74\times10^{-4}$, a factor of 51. The absolute information remains small, so this illustrates local gain through ejecta information. Early spectral velocities can supply such a constraint when validated transfer links them to the same ejecta parameter.

Equation~(\ref{eq:gain}) ranks independent measurements by their response mismatch $q-r^T\hat x$, weighted by uncertainty and compensation covariance. The greatest gain comes from a precise observable that responds differently to fission and its best compensation. Shared errors and extra spectral parameters enter the corresponding joint model.

Beyond ejecta constraints, additional photometric bands are informative through both counting precision and parameter dependent color responses. To separate these effects, consider $F_{ib}=p_{ib}L_i(\theta,\eta)$ with known fractions $p_{ib}$ independent of the fitted parameters. Here $i$ indexes epochs, $b$ indexes bands, and $F_{ib}$ is band flux in a common luminosity to flux normalization. The log Jacobian rows then repeat, preserving parameter rank. For two equal error bands with fractional uncertainty $\sigma$ and same epoch correlation $\rho$, the common luminosity information is $2/(1+\rho)$ times that of one band: at $\rho=0$ precision improves by $\sqrt2$, while at $\rho=1$ the measurements repeat the same information. Parameter dependent colors can instead constrain composition or ejecta and lift a projected degeneracy. Their physical responses involve ionization, line opacity, heating and viewing angle \cite{Bulla2023,Pognan2025}. Elemental emission lines supply another composition diagnostic, whose connection to a particular radioactive parent requires a validated source to spectrum calculation. Wavelength dependent, ideally non-LTE transfer and the cross band covariance define the joint model needed to quantify that gain \cite{Gillanders2025,Pognan2025}.

Coupled network and atomic opacity calculations demonstrate how these inputs change ejecta evolution and colors \cite{Magistrelli2026}. Similar normalized ionization efficiencies for alpha, beta and fission channels in examined compositions further show why source labels alone do not determine their spectral distinction \cite{VandenBerg2026}.

Correlated yields propagated through deposition and spectral transfer can therefore be evaluated in the same source deletion likelihood. Ejecta measurements and gravitational wave constraints, connected through merger modelling, restrict the competing responses.

\subsection{Energy conserving frequency redistribution}\label{sec:twogroup}
Bolometric summation discards frequency information. We introduce an energy conserving shell extension with higher and lower opacity radiation groups $B$ and $R$. Each source injects half its deposited energy into each group; unequal escape and a common redistribution law convert different heating histories into different emerging fractions. Within each shell,
\begin{equation}
\begin{aligned}
 \dot E_B&=\tfrac12Q_s-E_B/t-L_B-J,\\
 \dot E_R&=\tfrac12Q_s-E_R/t-L_R+J,\\
 J&=\frac{0.5}{t}\big[(1-\chi)E_B-\chi E_R\big].
\end{aligned}\label{eq:twogroup}
\end{equation}
Here $Q_s=M_s\qth$, $\chi(t)=0.3/[1+(t/30\,\mathrm d)^{1.5}]$, and the positive transfer rates conserve radiation energy between groups. Their escape luminosities use Eq.~(\ref{eq:transport}), with opacity factors $3g(t)$ and $0.3g(t)$ multiplying each shell's baseline opacity, where $g(t)=1+[1+(t/30\,\mathrm d)^2]^{-1}$. These specified factors introduce a frequency contrast and time dependent transparency with the same injection fractions for every source. The emerging group fractions also depend on each source history, so common injection fractions allow distinct output colors. They define a controlled non-gray calculation; atomic line formation and NLTE calibration remain the role of a wavelength-dependent transfer model \cite{Bulla2023,Pognan2025}.

Summing Eq.~(\ref{eq:twogroup}) cancels $J$ and preserves the finite-window energy identity. Equalizing both opacities to the original gray value recovers its total luminosity within $2.1\times10^{-15}$ over integer days 5--90. Refining 120 shells and 0.015 d steps to 200 shells and 0.005 d steps changes group luminosities by at most 0.0224\%.

The extra information can be separated exactly from flux precision. At epoch $i$, let $T_i=T_{Bi}+T_{Ri}$, $w_{Bi}=T_{Bi}/T_i$ and $w_{Ri}=1-w_{Bi}$. Assign independent group variances $\sigma_i^2w_{Bi}$ and $\sigma_i^2w_{Ri}$, with $\sigma_i=0.1T_i$. Their summed variance is precisely the original total-flux variance. For group residuals $r_B=T_B-N_B$ and $r_R=T_R-N_R$, suppressing the epoch index,
\begin{equation}
\begin{aligned}
 \frac{r_B^2}{\sigma^2w_B}+\frac{r_R^2}{\sigma^2w_R}
 ={}&\frac{(r_B+r_R)^2}{\sigma^2}\\
 &+\frac{(w_Rr_B-w_Br_R)^2}{\sigma^2w_Bw_R}.
\end{aligned}\label{eq:coloridentity}
\end{equation}
The first term measures total-flux separation; the nonnegative second term measures a color-response mismatch. If the predicted group fractions are independent of the fitted sources and ejecta, that mismatch vanishes and splitting the flux adds no information at this normalization. Otherwise the best joint fit has at least the best bolometric distance, because Eq.~(\ref{eq:coloridentity}) holds for every admissible alternative.

The frequency dependent opacity and redistribution reshape the total emission as well as its division between groups. At the unchanged injected alpha and Cf yields, the two-group truth retains a 38.03\% day 90 Cf share and gives a bolometric distance $d_G^2=0.05725$, compared with 0.08996 for the single-gray calculation. Resolving both groups then raises the distance to 0.47388 at the same total-flux precision within this two-group model.

\begin{figure*}[t]\centering
\includegraphics[width=0.485\textwidth]{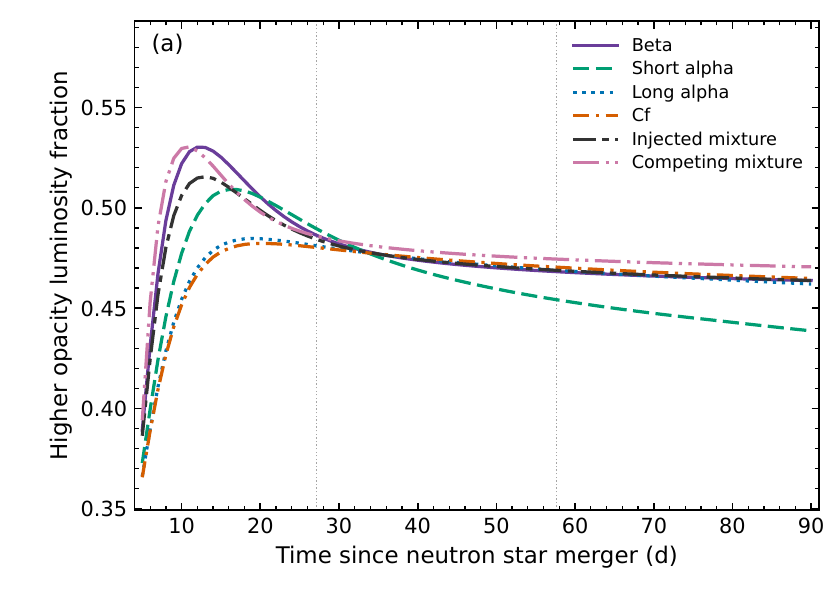}\hfill\includegraphics[width=0.485\textwidth]{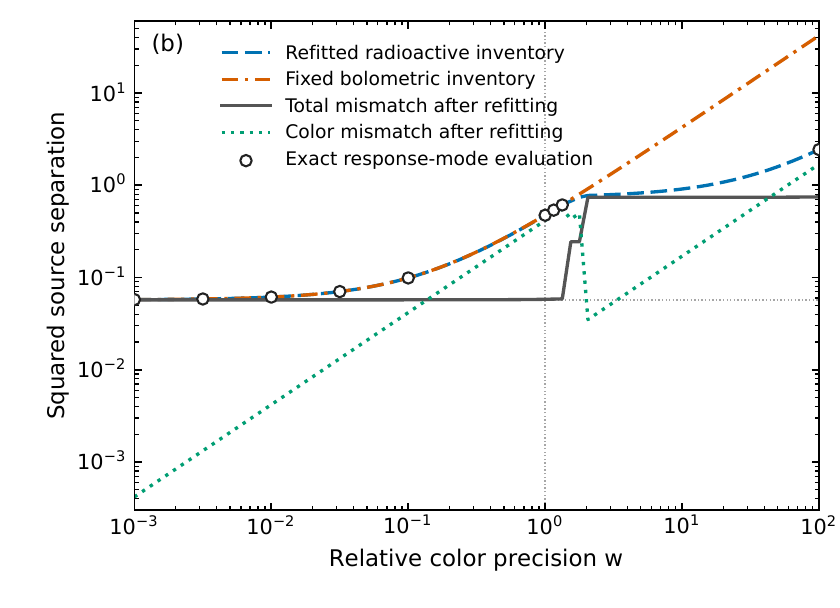}
\caption{Color increases separation while the preferred inventory readjusts. (a) Higher-opacity fraction $p_j=L_{Bj}/(L_{Bj}+L_{Rj})$ for each source at injected ejecta and for the injected and preferred competing mixtures. Vertical timing guides convert the AT~2023vfi JWST epochs from 28.9 and 61.4 observer days to 27.1 and 57.7 host days at $z=0.0647$ \cite{Gillanders2025}. These are calculated group fractions. (b) Optimized $\mathcal F(w)$, total and weighted color contributions, and the fixed bolometric-inventory upper bound. Guides mark $w=1$ and the bolometric minimum. Open circles reproduce exact response-mode predictions on verified active faces; independent face enumeration checks $w=1,100$. Total luminosity variance stays fixed; the groups provide a controlled radiation response rather than calibrated filters.}\label{fig:groups}\end{figure*}

At the joint optimum, total and color terms 0.05815 and 0.41573 sum to the joint distance within $10^{-9}$. The Gaussian ceiling is 16.94\% at 5\% false positives. Figure~\ref{fig:groups} resolves this additional color information and its precision dependence.

Color constrains the full compensating inventory. Long alpha and Cf remain closely aligned (cosine approximately 0.987 in total and whitened two-group responses), with emerging fractions differing by at most 0.329 percentage points. Enforcing common emerging fractions returns the joint distance to the bolometric value 0.057252. At the preferred alternative, $M=0.023\Msun$, $v_{\min}=0.08c$ and opacity multiplier 0.5; these changes also reshape beta radiation. Decompose the whitened color residual as $c=\sum_jc_j$ over beta normalization, beta tilt, short alpha, long alpha and deleted Cf. Then $Q_C=\|c\|^2=\sum_jc_j^Tc$. The beta-background/ejecta term contributes 0.3230 of $Q_C=0.4157$, or 77.7\%, in this signed decomposition. Its projections include cross terms and are not independent information fractions. Figure~\ref{fig:groups}(a) shows the source and mixture fractions; calibrated spectral transfer supplies their counterparts in observed filters.

\subsection{Source separation as a function of color precision}\label{sec:precision}
We derive a variational relation that quantifies how a competing nuclear inventory absorbs improved color information. Let $Q_T(e,x)$ and $Q_C(e,x)$ be the sums of the first and second terms of Eq.~(\ref{eq:coloridentity}). Introduce a relative color precision $w\geq0$, while retaining the same total luminosity variance, and define
\begin{equation}
 \mathcal F(w)=\min_{e\in G,\,x\in\mathcal P}
       \{Q_T(e,x)+wQ_C(e,x)\}.
 \label{eq:colorprofile}
\end{equation}
Thus $\mathcal F(0)$ is the bolometric distance and $\mathcal F(1)$ is the two group distance above. For $w>0$, this describes a Gaussian experiment in the fixed total and color coordinates $y_T=y_B+y_R$ and $y_C=w_Ry_B-w_By_R$, with independent variances $\sigma^2$ and $\sigma^2w_Bw_R/w$ at each epoch. The fractions $w_B,w_R$ are held at the stipulated truth for every candidate mean. At $w=1$ the original independent group variances are recovered. Varying $w$ changes color precision alone, rather than introducing additional total flux precision.

For any minimizing inventory at $w$, denote its residual sums by $Q_T(w)$ and $Q_C(w)$. Since Eq.~(\ref{eq:colorprofile}) is a lower envelope of affine functions of $w$, it is nondecreasing and concave \cite{Boyd2004}. More explicitly, evaluating either optimum at the other weight gives, for $w_2>w_1$,
\begin{equation}
 Q_C(w_2)\leq
 \frac{\mathcal F(w_2)-\mathcal F(w_1)}{w_2-w_1}
 \leq Q_C(w_1).
 \label{eq:precisionlaw}
\end{equation}
These inequalities follow directly by adding the two optimality comparisons; they do not require a convex union of ejecta templates. They imply $Q_C(w_2)\leq Q_C(w_1)$ and $Q_T(w_2)\geq Q_T(w_1)$. Wherever $\mathcal F$ is differentiable, $\mathcal F'(w)=Q_C(w)$ and $Q_T(w)=\mathcal F(w)-w\mathcal F'(w)$. Consequently, improved color precision raises source separation while the preferred competing inventory trades a larger total luminosity mismatch for a smaller color mismatch. At a template switch, Eq.~(\ref{eq:precisionlaw}) remains valid even when the derivative changes discontinuously.

The bolometric optimum supplies a useful upper bound on this gain:
\begin{equation}
 \mathcal F(0)\leq\mathcal F(w)
 \leq\mathcal F(0)+wQ_C(0).
 \label{eq:colorupper}
\end{equation}
Holding the bolometric inventory fixed gives the upper straight line; refitting gives the lower envelope. This standard value function construction \cite{Boyd2004} quantifies how nuclear compensation limits the gain from resolving radiation.

We evaluate 82 weights over all 27 ejecta templates, including $w=0$, 1 and 100. Every sampled pair satisfies Eq.~(\ref{eq:precisionlaw}); the global primal--dual gaps remain below $1.5\times10^{-9}$. Independent enumeration of constraint faces at the three anchor weights agrees within $7\times10^{-13}$. The profile reproduces the bolometric and original two group distances. At $w=100$, corresponding to a tenfold reduction in color standard error at fixed total error, the optimized distance is 2.4401. Holding the bolometric alternative fixed would instead give 41.9957. The difference resolves the source readjustment directly: the optimized total mismatch rises to 0.7490, while the unweighted color mismatch falls from 0.4194 to 0.01691. Even this improved precision allows at most 46.70\% power at a 5\% false-positive level in the declared experiment. Thus frequency resolution contributes a distinct physical response, and its identifying power is governed jointly by that response and the nuclear inventories permitted to compensate it.

\subsection{Response modes and the residual color floor}\label{sec:modes}
The color precision profile can be resolved into the source adjustments that absorb the new radiation information. Within one ejecta template and one fixed face of the coefficient constraints, write the admissible displacement as $x=x_0+Vu$, where the columns of $V$ span the free directions and $x_0$ is stationary at weight $w_0$. Let $\mathsf R_T$ and $\mathsf R_C$ be the whitened total and color response matrices in those directions, $c_0$ the color residual at $x_0$, and $H_0=\mathsf R_T^T\mathsf R_T+w_0\mathsf R_C^T\mathsf R_C$. For an additional precision $s=w-w_0\geq0$, expansion about that stationary solution gives
\begin{equation}
 \mathcal F_{\rm face}(w_0+s)=\mathcal F_{\rm face}(w_0)
 +\min_u\{u^TH_0u+s\|c_0-\mathsf R_Cu\|^2\}.
 \label{eq:modequadratic}
\end{equation}
When $H_0$ is positive definite on the free directions, stationarity gives $u_s=s(H_0+s\mathsf R_C^T\mathsf R_C)^{-1}\mathsf R_C^Tc_0$. If $H_0=LL^T$ and $\lambda_k$ are the squared singular values of $\mathsf R_C L^{-T}$, resolve $c_0$ into its corresponding left singular directions with coefficients $g_k$ and an orthogonal remainder $c_\perp$. The exact face profile is
\begin{equation}
 \mathcal F_{\rm face}(w_0+s)=\mathcal F_{\rm face}(w_0)
 +s\|c_\perp\|^2+\sum_k\frac{s g_k^2}{1+s\lambda_k}.
 \label{eq:responsemodes}
\end{equation}
Each positive $\lambda_k$ identifies a color response that the permitted inventory adjustment can partly absorb. Its contribution bends away from the frozen-inventory straight line as precision improves. The orthogonal remainder instead accumulates separation linearly. The expression applies while the stationary coefficients remain on the same feasible face and satisfy its optimality conditions; the global profile selects among these branches and templates. It reproduces 1,175 anchor-to-weight evaluations of the existing profile within $1.0\times10^{-11}$ in squared distance. Thus the curvature in Fig.~\ref{fig:groups}(b) resolves the changing nuclear compensation, through the spectral form of quadratic projection \cite{Boyd2004}.

A global statement follows without fixing a face. Let $\mathcal C_*=\min_{e,x}Q_C(e,x)$ and let $(e_C,x_C)$ be any attaining color-only fit, with $T_C=Q_T(e_C,x_C)$. Compactness of the bounded family gives
\begin{equation}
 \mathcal F(0)+w\mathcal C_*\leq\mathcal F(w)\leq T_C+w\mathcal C_*,
 \qquad\lim_{w\to\infty}\frac{\mathcal F(w)}{w}=\mathcal C_*.
 \label{eq:colorfloor}
\end{equation}
The certified color-only minimum $\mathcal C_*=7.6240\times10^{-4}$ over 27 templates is reproduced by constraint-face enumeration. Its positive value identifies contrast that the permitted two-group inventories cannot cancel. The color optimum reaches the beta amplitude and shape boundaries but has total mismatch $T_C=106.77$, far above the bolometric minimum. Color and total luminosity therefore favor different allocations; increased color precision progressively shifts the preferred compensation. This floor quantifies retained color contrast for the stated family and radiation responses. Composition-sensitive lines can add independent response directions, while freedom in ionization, line opacity and composition can supply additional compensation. Wavelength-dependent transfer determines their balance and maps the calculated groups to measured passbands \cite{Pognan2025,Gillanders2025}. The groups represent opacity ranges; a comparison with JWST or Hubble filters follows by integrating calibrated spectra over each instrumental passband.

\section{Astrophysical implications}
\subsection{A decline benchmark from late infrared observations}
Late JWST and Hubble Space Telescope observations of AT~2023vfi connect the response comparisons to measured radiation. Its thermal reconstruction after separating the afterglow contribution gives $L_{\rm bol}\propto t^{-\nu_{\rm obs}}$, with $\nu_{\rm obs}=2.7\pm0.4$ \cite{Yang2024}. We use this reported decline and uncertainty as a one-statistic benchmark over the JWST interval.

\begin{figure*}[t]\centering
\includegraphics[width=0.485\textwidth]{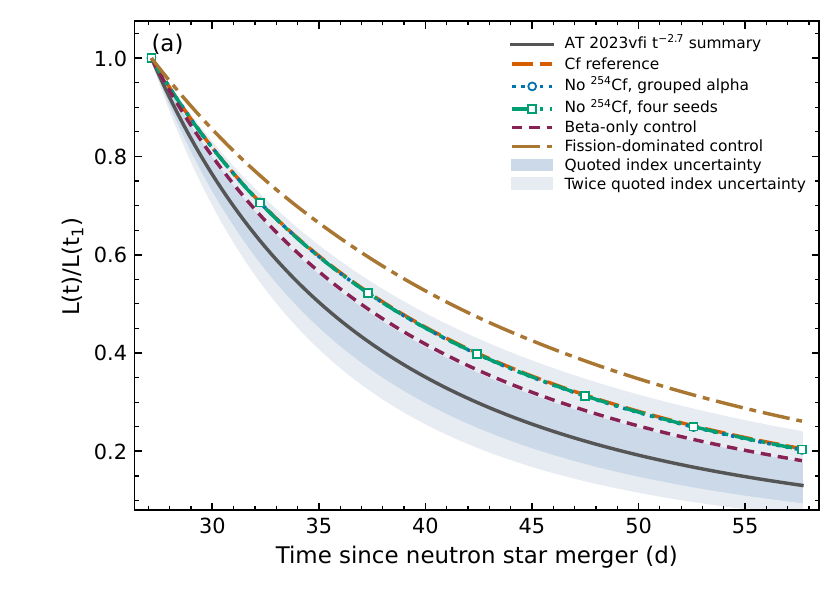}\hfill\includegraphics[width=0.485\textwidth]{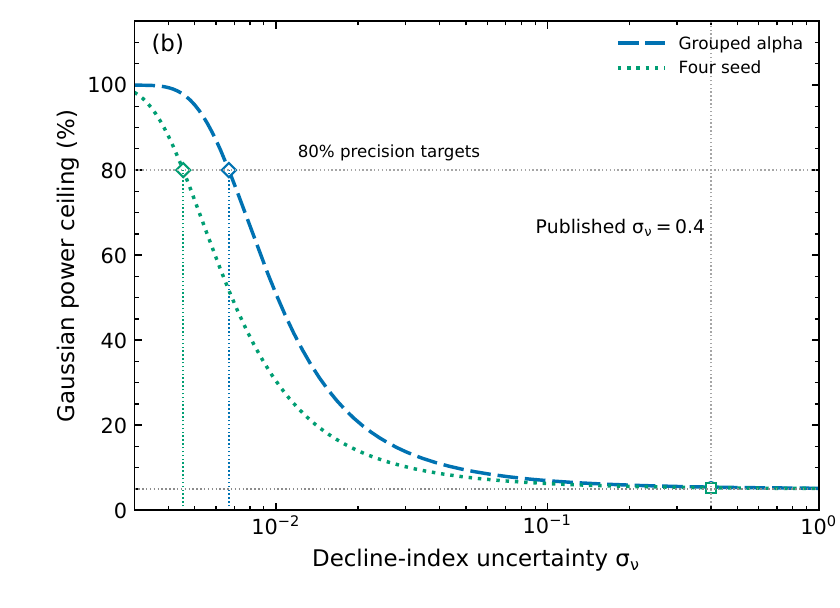}
\caption{Observed decline precision and the range of transported source mixtures. (a) Cf, retained no-Cf, beta-only and constructed fission-dominated curves normalized at the first JWST epoch. The latter increases $Y_{254}$ tenfold at reference ejecta. The central AT~2023vfi summary and bands use index $2.7\pm0.4$ \cite{Yang2024}; the outer band doubles that uncertainty. The retained reference and alternatives predict declines about 1.5 quoted uncertainties shallower than the reconstruction and differ much less from each other. These bands represent correlated index uncertainty at fixed normalization. (b) Gaussian power ceilings for the retained pairs versus decline-index uncertainty. Open symbols mark the published uncertainty; guides show 5\% false positives and the 80\% precision targets. Both panels describe the summary benchmark, with no event fit.}\label{fig:decline}\end{figure*}

For a calculated light curve define
\begin{equation}
 \nu_{12}[L]=-\frac{\ln[L(t_2)/L(t_1)]}{\ln(t_2/t_1)},\qquad
 t_{1,2}=\frac{(28.9,61.4)\,\mathrm d}{1+0.0647}.
 \label{eq:decline}
\end{equation}
The ratio removes a common normalization. Direct integration gives $\nu_{12}=2.101$ for the Cf reference and 2.117 and 2.112 for its existing grouped and four-seed alternatives. These retained eighteen-epoch solutions predict declines about $1.5$ quoted uncertainties shallower than the reconstructed decline, and their mutual differences remain within the reported decline uncertainty (Fig.~\ref{fig:decline}). This comparison evaluates their normalized temporal response without fitting the event.

For the illustrative summary likelihood $\widehat\nu\sim\mathcal N(\nu_{12},\sigma_\nu^2)$, $\sigma_\nu=0.4$, the retained witnesses give $d_\nu=|\nu_{12,T}-\nu_{12,N}|/\sigma_\nu$. Equation~(\ref{eq:powerbound}) gives 5.44\% and 5.30\% ceilings at 5\% false positives. More directly, an 80\% fixed-pair power ceiling requires $\sigma_\nu\leq|\Delta\nu_{12}|/[z_{0.95}+z_{0.80}]$: 0.00669 and 0.00454 for these alternatives, approximately 60 and 88 times tighter than the published uncertainty. These necessary pairwise targets quantify the gain required from decline measurements; source profiling determines the corresponding requirement for the full family. The Gaussian approximation uses the quoted reconstruction uncertainty as a summary scale, retaining its dependence on the observational modelling.

Fixed-transport source controls reveal the wider temporal range. At the reference ejecta, beta-only, reference-proportion alpha-only and Cf-only luminosities give $\nu_{12}=2.268$, 2.371 and 1.663. Increasing only the reference Cf abundance tenfold to $Y_{254}=2\times10^{-5}$ gives 1.781, with Cf supplying 77.5\% and 84.7\% of the luminosity at the interval endpoints. This constructed fission-dominated control isolates source strength at fixed transport. Its index and the Cf-only index lie approximately 2.3 and 2.6 quoted uncertainties below the reconstructed central value, giving larger offsets than the compensated inventories within the same gray response model. The three pure-channel controls span 0.708 in index, compared with only 0.011--0.017 between the compensated inventories. Thus a decline can distinguish widely separated source mixtures more readily than inventories whose full responses compensate.

Linearity gives this statement an exact form. For $L=\sum_j a_jL_j$ with $a_j\geq0$, let $w_j=a_jL_j(t_1)/L(t_1)$ and $\ell=\ln(t_2/t_1)$. Then
\begin{equation}
 \nu_{12}[L]=-\frac{1}{\ell}\ln\!\left[\sum_j w_j e^{-\ell\nu_{12}[L_j]}\right],\qquad \sum_jw_j=1.
 \label{eq:mixturedecline}
\end{equation}
The mixture index lies between the contributing source indices. For beta and Cf alone it decreases monotonically with the Cf fraction at $t_1$, because Cf has the shallower transported decline. Source-dependent deposition and changes in ejecta alter the response indices themselves, supplying the additional freedom used by the fitted alternatives. A 200-shell, 0.005 d integration reproduces the controls within $0.008\%$ in luminosity and $10^{-5}$ in index.

The bands in Fig.~\ref{fig:decline}(a) propagate one reconstructed index uncertainty at fixed normalization. Yang et al. connect the rapid decline and photospheric recession to lanthanide recombination \cite{Yang2024}; the present gray response holds the opacity prescription fixed. In their analysis of the kilonova bolometric luminosity, Yang et al. show that matching the weak 61 d emission calls for a lower $^{254}$Cf yield or reduced deposition of its decay energy \cite{Yang2024}. The observational interpretation and source controls motivate joint constraints on source yields, deposition and evolving opacity. Absolute luminosity and spectra supply the complementary information for an event likelihood.

\subsection{From response comparisons to event inference}
The timing anchors in Fig.~\ref{fig:groups}(a) convert the 28.9 and 61.4 observer-frame JWST days to 27.1 and 57.7 host-frame days at $z=0.0647$ \cite{Gillanders2025}; the illustrative 29 and 61 d sampling design uses model-time epochs motivated by that cadence. The retained inventories predict declines about 1.5 quoted uncertainties shallower than the reconstructed observable. The source controls resolve a wider range of temporal responses, while the total/color decomposition identifies complementary spectral information.

The same attribution formulation can be applied to radioactive electromagnetic counterparts of disruptive neutron star--black hole mergers by supplying their source inventories and ejecta radiation responses \cite{Metzger2020}.

For an individual event, correlated yields resolved by electron fraction determine which radioactive allocations are available. Wavelength dependent opacity and multidimensional transfer then determine how those allocations appear in the measured radiation. Independent ejecta constraints and gravitational wave binary information, connected through merger modelling, restrict the same compensating family. Distance, background and cross epoch covariance specify the likelihood in place of the independent 10\% reference errors. The matched luminosity controls, bounded source deletion and color response modes provide a common comparison across these richer models \cite{Barnes2016,Bulla2023,Pognan2025,Magistrelli2026}. Their role is to identify which additional physical constraint separates a radioactive parent from the inventories that already explain its bolometric contribution.

\section{Conclusion}
Radioactive power and unique source identification probe distinct properties of neutron star merger emission. This study develops a unified model that connects nuclear heating, particle deposition and radiation transport to quantitative conditions for isotope attribution. Matched luminosity comparisons show how deposition aligns long alpha and fission responses beyond changes in their amplitudes. Explicit beta and alpha inventories reproduce light curves containing substantial late $^{254}$Cf power, revealing the source and ejecta adjustments that preserve the radiation signature after fission is removed.

Optimization certificates verify the bounded source comparisons, while Gaussian likelihood bounds translate their separation into limits on identification. The collective bound strengthens these limits by incorporating several admissible inventories and resolving the roles of their mean displacement and dispersion. For independently allocated alpha seeds, the analytic ceiling is 6.20\% at 5\% false positives despite a 38\% day-90 Cf contribution. Fixed-beta comparisons retain compensation through alpha allocation and ejecta adjustment, while the two-component calculations identify smooth evolution of the beta luminosity response as an important additional control. Nuclear yield restrictions and temporal sampling therefore determine which compensating inventories an isotope claim must distinguish.

Color supplies complementary information whose value depends on the permitted source adjustments. The energy conserving two-frequency-group model isolates this information at fixed total luminosity precision. Its nondecreasing, concave separation profile describes how the preferred competing inventory readjusts as color precision improves. Exact response modes resolve the adjustments that absorb contrast, and a positive residual floor quantifies the contrast retained by the specified family. The source decomposition identifies the beta background and accompanying ejecta changes as major contributors to the color residual, connecting spectral discrimination to the complete radioactive inventory.

Late observations of AT~2023vfi place these results on an observational scale. The reference and competing no-Cf inventories predict bolometric declines about 1.5 quoted uncertainties shallower than the reconstruction from James Webb and Hubble Space Telescope observations, while their mutual differences remain within the reported uncertainty. Fixed-transport controls reveal a wider range of temporal responses, with fission-dominated mixtures producing shallower declines and larger offsets from the reconstructed value. The exact mixture relation explains how source allocation shapes this behavior. The observational interpretation also highlights the contribution of recombination and evolving opacity, motivating joint use of luminosity, spectral evolution and calibrated radiation transport.

This work provides a quantitative foundation for connecting nuclear production, particle deposition and radiation transport to observable merger signatures. Bounded source comparisons and response geometry identify how nuclear yield constraints, cadence and color precision sharpen radioactive attribution. These results inform the design and interpretation of merger observations and the use of electromagnetic and gravitational wave constraints to determine which radioactive nuclei can be identified in kilonovae.

\end{document}